\documentclass[useAMS,usenatbib]{mn2e}
\usepackage{graphicx}
\usepackage{latexsym}
\usepackage{psfig}
\usepackage{amssymb,amsmath}
\usepackage{subfigure}

\title[Comparative mass segregation]{Comparisons between different techniques for measuring mass segregation}

\author[R. J. Parker \& S. P. Goodwin]{Richard  J. Parker$^{1}$\thanks{E-mail: R.J.Parker@ljmu.ac.uk} and Simon P. Goodwin$^{2}$  \vspace*{0.1cm}\\
$^{1}$Astrophysics Research Institute, Liverpool John Moores University, 146 Brownlow Hill, Liverpool, L3 5RF, UK\\
$^{2}$Department of Physics and Astronomy, University of Sheffield, Sheffield, S3 7RH, UK}

\begin{document}

\pagerange{\pageref{firstpage}--\pageref{lastpage}} \pubyear{2015}

\maketitle

\label{firstpage}

\def\mnras{MNRAS}
\def\apj{ApJ}
\def\aj{AJ}
\def\aap{A\&A}
\def\apjl{ApJL}
\def\apjs{ApJS}
\def\araa{ARA\&A}
\def\pasj{PASJ}
 
\begin{abstract}
We examine the performance of four different methods which are used to
measure mass segregation in star-forming regions: the radial variation
of the mass function $\mathcal{M}_{\rm MF}$; the minimum spanning
tree-based $\Lambda_{\rm MSR}$ method; the local surface density
$\Sigma_{\rm LDR}$ method; and the $\Omega_{\rm GSR}$ technique, which
isolates groups of stars and determines whether the most massive star
in each group is more centrally concentrated than the average star. 
  All four methods have been proposed in the literature as techniques
  for quantifying mass segregation, yet they routinely produce
  contradictory results as they do not all measure the same thing. We apply each method to synthetic star-forming regions to determine when and why they have shortcomings.  When a star-forming region is smooth and centrally concentrated, all four methods correctly identify mass segregation when it is present. However, if the region is spatially substructured, the $\Omega_{\rm GSR}$ method fails because it arbitrarily defines groups in the hierarchical distribution, and usually discards positional information for many of the most massive stars in the region. We also show that the  $\Lambda_{\rm MSR}$ and $\Sigma_{\rm LDR}$ methods can sometimes produce apparently contradictory results, because they use different definitions of mass segregation. We conclude that only $\Lambda_{\rm MSR}$ measures mass segregation in the classical sense (without the need for defining the centre of the region), although $\Sigma_{\rm LDR}$ does place limits on the amount of previous dynamical evolution in a star-forming region.    
\end{abstract}


\section{Introduction}

Most young ($<$10\,Myr) stars are observed in the company of others,
either in clusters \citep{Lada03}, groups \citep{Porras03},
associations \citep{Blaauw64} or in regions which are over-dense with
respect to the Galactic field \citep{Bressert10}. Attempts have been
made at distinguishing isolated versus clustered star-formation, or
clusters versus associations
\citep[e.g.][]{Cartwright04,Gieles11}. However, it appears that (at
least on local scales) there is no fundamental scale unit for star
formation \citep[e.g.][]{Bressert10}, and that clusters may just be
the dense tail of the distribution of Galactic star formation
\citep{Kruijssen12b}. 

Given that star formation on pc-scales often seems to produce complex,
hierarchical distributions, it is crucial to be able to examine the spatial
distributions of stars in a quantitative, statistical way.  This
allows us to extract information on star formation and compare and
contrast different regions (and simulations).  One such indicator is the relative
positions of the most massive stars with respect to the average stars
in a region -- often referred to as `mass segregation' when the most
massive stars are more centrally concentrated than average. 

In recent years several alternative methods for measuring
mass segregation have been proposed in the literature. 
In this paper we critically assess several of the new methods, including the
$\Lambda_{\rm MSR}$ method \citep{Allison09a}, the $\Sigma - m$ or
$\Sigma_{\rm LDR}$ method \citep{Maschberger11} and the technique of
dividing a star-forming region into groups to determine the relative
distance of the most massive star to the centre of each group
\citep{Kirk10,Kirk14}, which we call $\Omega_{\rm GSR}$. A similar
study to compare mass segregation methods was previously carried out
by \citet{Olczak11}; however, the $\Sigma_{\rm LDR}$ and $\Omega_{\rm
  GSR}$ methods were not yet prominent in the literature and therefore
not included in that study. A full appraisal of these methods -- and
$\Lambda_{\rm MSR}$ -- using synthetic data is therefore 
required in order to make sense of the growing literature in this
area.

The paper is organised as follows. In Section~\ref{mass_seg:define} we
provide a brief summary of the different methods used to define mass
segregation, and discuss the methods we will use in this paper in
detail. In Section~\ref{synth:data} we test these methods on synthetic
data, before discussing our results in Section~\ref{discuss} and
concluding in Section~\ref{conclude}.

\section{What is mass segregation?}
\label{mass_seg:define}

`Mass segregation' is a phrase often used in relation to star clusters
and regions, but often it is not defined.

The classical definition of mass segregation is based on the behaviour
of dynamically old bound systems (i.e. relaxed and virialised star
clusters).  The process of two-body relaxation redistributes energy
between stars and they approach energy equipartition whereby all stars
have the same mean kinetic energy -- this means that more massive
stars will have a velocity dispersion that is lower.  Because the velocity
dispersion of the more massive stars is lower, they will tend to be
concentrated towards the centre of the cluster \citep{Spitzer69}.

The timescale, $t_{\rm MS}(M)$, on which stars of mass $M$ will dynamically mass
segregate depends on the mean mass of stars $\left<m\right>$ and the two-body
relaxation time, $t_{\rm relax}$ \citep{Spitzer69}
\begin{equation}
t_{\rm MS}(M) \sim \frac{M}{\left<m\right>} t_{\rm relax}.
\end{equation}
Therefore in old star clusters (especially globular clusters) we
expect to see mass segregation down to low masses and that mass
segregation is explicable entirely by dynamics.

In these old clusters, the two-body dynamics that may have caused any mass segregation will also remove any primordial substructure in the spatial distribution and we would expect (and observe) the cluster to have a smooth, centrally concentrated spatial distribution, such as a \citet{Plummer11} or \citet{King66} profile. 

In this case, the clusters have a well-defined radial profile where one can quantify mass segregation by taking different mass bins and comparing the density profiles \citep[e.g.][]{Hillenbrand97,Pinfield98}, or examining variations in the slope of the mass function (or luminosity function) with distance
from the cluster centre \citep{Carpenter97,deGrijs02,Gouliermis04}. A related method is to quantify the variation of the `Spitzer radius' -- the rms distance of stars in a cluster around the centre of mass -- with luminosity \citep{Gouliermis09}.

The motivation in attempting to observe mass segregation in young
clusters or regions is that it might not have a dynamical origin.  If
we observe mass segregation in a region that is so young 
that two-body encounters cannot have mass segregated the stars\footnote{This is complicated by the fact that it is not the current
relaxation time of the region that is important, rather it is how
dynamically old the region is \citep[see][]{Allison10}.}, then
the mass segregation must be set by some aspect of the star formation
process, and is often labelled `primordial mass segregation'. 

Primordial mass segregation has been found in some simulations of star
formation \citep[e.g.][]{Maschberger11,Myers14} but not in others
\citep{Girichidis12,Parker15a}, and we also note that any observed
signature may be a combination of primordial and dynamical mass
segregation \citep{Moeckel09a}, and
to complicate matters further mass segregation can be introduced
very rapidly dynamically through violent relaxation rather than
two-body relaxation \citep[see][]{Allison10}.

Observationally, mass segregation has been searched for in clusters and star forming
regions for many years \citep[e.g.][and many more]{Sagar88,deMarchi96,Hillenbrand98,Raboud98,deGrijs02,Gouliermis04,Sabbi08,Gouliermis09}, but in the past few years a number of new
statistical methods have been developed for finding mass segregation,
and the purpose of this paper is to examine what {\em exactly} they
are searching for, and what problems they might have.

Here we briefly describe the most commonly used methods and their
assumptions.  In the next section we describe exactly how we implement
each method in detail.

It should be noted that all methods suffer from a potentially very
serious problem.  All methods examine the relative distributions of
high- and low-mass stars.  Therefore the distribution of low-mass
stars must be known, however these are very faint compared to the
high-mass stars in young regions/clusters. And so the location in which the
observer is biased against observing low-mass stars is near luminous
high-mass stars which is exactly where one needs to know if low-mass
stars are present or not.  In this paper we use fake data in which we
have the advantage of knowing exactly where every star is and what its
mass is, but real observations do not have this advantage
\citep[see][for a detailed discussion of observational 
selection effects and biases]{Stolte05,Ascenso09}.

\subsection{Radial mass functions, $\mathcal{M}_{\rm MF}$}

Until recently, the most commonly used way of determining mass
segregation was to compare the radial distribution of the mass
function, or the radial distribution of stars of different masses \citep[e.g.][]{Sagar88,Gouliermis04,Stolte06,Sabbi08,Chavarria10}.  In this definition, if
a cluster is mass segregated the most massive stars are preferentially
towards the centre, and low-mass stars preferentially in the
outskirts.  Therefore the slope of the mass function should be flatter
in the centre than the outskirts, and/or the low-mass stars should
have a much broader radial distribution.

This method suffers three significant drawbacks.  One is that it
requires a `centre' to be defined.  In relaxed, virialised star
clusters there is a centre from which this can be measured, but in
substructured and messy young star forming regions it is unclear if
defining a `centre' means anything at all, and even if it did, it is
unclear how one would practically do this. One can define a geometric centre based on the positions of stars, but in some morphologies (as we shall see) this `centre' is not near to, nor is the central location, of the stars. 

Second, the definition of radial bins is non-trivial and often arbitrary, which  adds a further level of complexity to the interpretation of any signal.

It also suffers from poor statistics, in that it is unclear how to
compare different clusters or provide any {\em quantitative} information on
the mass segregation.  

In practice, we compare the cumulative distribution of the radii
of the ten most massive stars with the cumulative distribution of the
radii of all the stars in the distribution. We set the
`centre' of the region to be the (known to us) centre of mass of the
region.  We gauge
the significance of any difference using a Kolmogorov-Smirnov (KS)
test, and rather generously use a $p$-value of $<$\,0.1 as our significance threshold.

\subsection{The $\Lambda_{\rm MSR}$-parameter}

In order to try and avoid the problems of defining a centre and to
produce a quantitative statistic to aid comparisons Allison et
al. (2009) introduced the $\Lambda_{\rm MSR}$-parameter.

The $\Lambda_{\rm MSR}$-parameter examines the relative (2D) spatial
distributions of the $N$ most massive stars relative to each-other with the
spatial distributions of $N$ random stars.  This is done by finding
the edge length of the minimum spanning tree (MST) that connects the $N$
most massive stars with many MSTs of $N$ random stars.

This has the great advantage that it does not require a centre or any
special position to be defined.  It also produces a number with
associated error that states how much longer or shorter the massive star
MST is compared to random MSTs, and how likely it is that this length
could be drawn at random from the random MSTs (i.e. how likely is this
to occur by random chance).

The $\Lambda_{\rm MSR}$-parameter measures a very
similar `mass segregation' to the classical definition: it examines
if the massive stars are closer to each-other than one would expect by
random chance. If the massive stars are closer to one another, then the star-forming region is said to be mass segregated. 

It practice, we take the ratio of the average (mean) random MST length to the
subset MST length, a quantitative measure of the degree of  mass
segregation (normal or inverse) can be obtained. We first determine
the subset MST length, $l_{\rm subset}$. We then  determine the
average length of sets of $N_{\rm MST}$ random stars each time,
$\langle l_{\rm average} \rangle$. There is a dispersion  associated
with the average length of random MSTs, which is roughly Gaussian and
can be quantified as the standard deviation  of the lengths  $\langle
l_{\rm average} \rangle \pm \sigma_{\rm average}$. However, we
conservatively estimate the lower (upper) uncertainty  as the MST
length which lies 1/6 (5/6) of the way through an ordered list of all
the random lengths (corresponding to a 66 per cent deviation from  the
median value, $\langle l_{\rm average} \rangle$). This determination
prevents a single outlying object from heavily influencing the
uncertainty.  We can now define the `mass  segregation ratio'
($\Lambda_{\rm MSR}$) as the ratio between the average random MST
pathlength  and that of a chosen subset, or mass range of objects:
\begin{equation}
\Lambda_{\rm MSR} = {\frac{\langle l_{\rm average} \rangle}{l_{\rm
      subset}}} ^{+ {\sigma_{\rm 5/6}}/{l_{\rm subset}}}_{-
  {\sigma_{\rm 1/6}}/{l_{\rm subset}}}.
\end{equation}
A $\Lambda_{\rm MSR}$ of $\sim 1$ shows that the stars in the chosen
subset are distributed in the same way as all the other  stars,
whereas $\Lambda_{\rm MSR} > 1$ indicates mass segregation and
$\Lambda_{\rm MSR} < 1$ indicates inverse mass segregation,
i.e.\,\,the chosen subset is more widely distributed than the other
stars.

Note that a slightly different formulation of $\Lambda_{\rm MSR}$, which uses the geometric mean to define the uncertainties, is available \citep{Olczak11}, but for the purposes of this paper the original method by \citet{Allison09a} is sufficient.

\subsection{The local density ratio, $\Sigma_{\rm LDR}$}

\citet{Maschberger11} introduced another measure of mass segregation
-- the local density ratio.  For this the local surface density of
every star $\Sigma$ is found, and the average local surface density of the $N$
most massive stars is compared to the average local surface density of
all stars to obtain the `local surface density ratio', $\Sigma_{\rm LDR}$ \citep{Kupper11,Parker14b}.  This is able to determine if the most massive stars are in
regions of significantly higher (or lower) local surface density than
would be expected by random chance. If the most massive stars are in areas of higher local density than the region average, the region is said to be mass segregated.

As with $\Lambda_{\rm MSR}$, $\Sigma_{\rm LDR}$ makes no assumptions about there being a centre. 
However it is possible that the same numerical value of $\Sigma_{\rm LDR}$ can
be statistically significant sometimes, but not other times, which we quantify by means of a KS test.

It is very important to note that $\Lambda_{\rm MSR}$ and $\Sigma_{\rm LDR}$ measure
different versions of `mass segregation'.  $\Lambda_{\rm MSR}$ determines if the
massive stars are closer to each other than one would expect by random
chance, $\Sigma_{\rm LDR}$ determines if the massive stars are in regions of
higher surface density than one would expect by random chance.

It would be quite possible for $\Sigma_{\rm LDR}$ to find `mass segregation',
but for $\Lambda_{\rm MSR}$ not to.  This would occur for example if the massive
stars were widely distributed, but each had a local overdensity of
low-mass stars \citep[see][for examples]{Parker14b}.  

In practice, we calculate the local stellar surface density following the
prescription of \citet{Casertano85}, modified to account for the
analysis in projection. For an individual star the local stellar
surface density is given by
\begin{equation}
\Sigma = \frac{n - 1} {\pi r_{n}^2},
\label{sigma}
\end{equation}
where $r_{n}$ is the distance to the $n^{\rm th}$ nearest neighbouring
star\footnote{Note that this $n$ does not need to have the same value as the $N$ used to define the subset of choice, although we adopt $n = 10, N = 10$ throughout this work.}.

\citet{Kupper11} and \citet{Parker14b} took the ratio of the median
surface density of a chosen subset (in this paper we will use the 10
most massive stars) to the median for all stars in a region to define
the local surface density ratio, $\Sigma_{\rm LDR}$:
\begin{equation}
\Sigma_{\rm LDR} =
\frac{\tilde{\Sigma}_\mathrm{subset}}{\tilde{\Sigma}_\mathrm{all}}.
\end{equation}
The  $\Sigma_{\rm LDR}$ ratio is then quoted with the $p$-value from
the KS test to gauge the significance of any deviation from the median
for all stars, again with a $p$-value $<$\,0.1 used as the boundary between the difference being significant or not.

\subsection{Group segregation ratio, $\Omega_{\rm GSR}$}
\label{method:gsr}

A further, alternative method for quantifying mass segregation was
recently suggested by  \citet{Kirk10} and \citet{Kirk14}.  In this
method stellar groups are identified and `isolated' from the total
distribution.  Each of these groups is then examined to see if the
most massive star it contains is closer to the centre of the group 
than the median distance of all stars in the group. If the most massive star is closer to the centre than the average star in the majority of the groups, the star-forming region is said to be mass segregated.

The `mass segregation' searched for in this method is different again
from both $\Lambda_{\rm MSR}$ and $\Sigma_{\rm LDR}$.  $\Omega_{\rm
  GSR}$ divides the region into groups and then examines each group
for evidence of internal mass segregation.  In this process many stars
in the region (possibly including some of the most massive) can be
excluded if they do not belong in a group.

First, an MST is constructed for the entire region
and a cumulative distribution of all MST branch lengths is then
made. Two power-law slopes are then fitted to the shortest lengths,
and the longest lengths, and the intersection of these slopes defines
the boundary of subclustering, $d_{\rm break}$ \citep{Gutermuth09}.

The links in the full MST which exceed  $d_{\rm break}$ are removed,
resulting in the star-forming region being divided into groups. If the
position of the most massive star in the group $r_{\rm mm}$ is closer
to the central position than the median value for all stars, $r_{\rm
  med}$, the group, or subcluster has an offset ratio ($r_{\rm
  mm}/r_{\rm med}$) less than unity and is said to be mass segregated.  

In this paper, we define a `group segregation ratio', $\Omega_{\rm
  GSR}$, as 
\begin{equation}
\Omega_{\rm GSR} = \frac{N_{\rm seg}}{N_{\rm grp}},
\end{equation}
where $N_{\rm grp}$ are the number of groups, and $N_{\rm seg}$ is the
number of these groups that have an offset ratio less than unity. If a
star-forming region has $\Omega_{\rm GSR} = 1$,  then all individual
groups defined by $d_{\rm break}$ are mass segregated. 

It should be noted that, just by random chance we would expect $\Omega_{\rm
  GSR} \sim 0.5$, as half of the time the most massive star would be in
the inner 50 per cent of stars.  The significance of $\Omega_{\rm
  GSR} \sim 0.5$ is affected by Poisson noise; for example
an $\Omega_{\rm GSR} = 0.8$ is not significant if it is 8 out
of 10 subgroups.

\section{Finding mass segregation in simulated regions}
\label{synth:data}

All of these four methods for finding `mass segregation' will find
classical mass segregation in a relaxed, spatially smooth, spherical, bound cluster.  If the most
massive stars are close together in the centre of a spherical cluster
then: (A) $\mathcal{M}_{\rm MF}$ will show a different mass function
in the inner regions.  (B) $\Lambda_{\rm MSR}$ will show that the
massive stars are concentrated together.  (C) $\Sigma_{\rm LDR}$ will
find that the most massive stars are in the regions of highest surface
density.  (D) $\Omega_{\rm GSR}$ will find that the most massive star 
is towards the centre of a single group (in this situation, the cluster itself is the group).

In such a situation we would advise the reader to use $\Lambda_{\rm
  MSR}$ to look for mass segregation as it gives a single 
quantitative value for the degree of mass segregation with an
associated error, and can easily determine the stellar mass down to which mass mass
segregation is present \citep[e.g.][]{Allison09a,Sana10,Beccari12,Delgado13,Er13,Pang13,Rivilla14,Wright14}.

What we will examine in this paper is the analysis of complex,
substructured regions and the search for `mass segregation' within
them, according to the definition presented in each method.

In this Section we run our four methods for quantifying mass
segregation on a synthetic dataset containing $N = 300$ stars to match
the small-$N$ statistics of many young regions. We
distribute the stars in a fractal distribution according to the
prescription in \citet{Goodwin04a}, \citet{Allison10} and
\citet{Parker14b}. We refer the reader to those papers for a detailed
description of how the fractal is set up, but we briefly summarise the
method here. The fractal is built by creating a cube containing
`parents', which spawn a number of `children' depending on the desired
fractal dimension, $D$. The amount of substructure is then set by the number of
children that are allowed to mature (the lower the fractal dimension,
the fewer children mature and the cube has more substructure). 

We choose a fractal distribution because the $\Lambda_{\rm MSR}$,
$\Sigma_{\rm LDR}$ and $\Omega_{\rm GSR}$ methods were developed
specifically to be used on substructured, or hierarchical spatial
distributions of stars in star-forming regions and clusters. All three
negate the requirement of designating a `centre' (although the
$\Omega_{\rm GSR}$  method requires the definition of group centres --
as detailed in Section~\ref{method:gsr}). Star-formation may result in
a truly fractal distribution \citep{Elmegreen01,Cartwright04}, but
star forming regions are unlikely to retain their primordial spatial
distribution due to dynamical evolution \citep{Parker14b}. As a
default, we choose $D = 2.0$ and assign the fractal a radius of 5\,pc.

We note that observed star-forming regions display a range of fractal dimensions \citep{Cartwright04,Schmeja08,Schmeja09,Sanchez09,Gouliermis14}. It is possible that all regions may form with the same (low) fractal dimension, and observed differences are due to differing amounts of dynamical evolution \citep{Goodwin04a,Parker14b,Parker14e}, although simulations of star-formation also produce a range of values before significant dynamical evolution takes place \citep{Girichidis12,Dale12a,Dale13}. For the purposes of the numerical tests presented here, regions with $D = 2.0$ adequately highlight the differences between the various definitions of mass segregation.

We draw masses from the \citet{Maschberger13} formulation of the
initial mass function (IMF):
\begin{equation}
p(m) \propto \left(\frac{m}{\mu}\right)^{-\alpha}\left(1 +
\left(\frac{m}{\mu}\right)^{1 - \alpha}\right)^{-\beta}
\label{imf}.
\end{equation}
Eq.~\ref{imf} essentially combines the log-normal approximation for
the IMF derived by \citet{Chabrier03,Chabrier05} with the
\citet{Salpeter55} power-law slope for stars with mass
$>$1\,M$_\odot$. Here, $\mu = 0.2$\,M$_\odot$ is the average stellar
mass, $\alpha = 2.3$ is the Salpeter power-law exponent for higher
mass stars, and $\beta = 1.4$ is the power-law exponent to describe
the slope of the IMF for low-mass objects \citep*[which also deviates
  from the  log-normal form;][]{Bastian10}. Finally, we sample from
this IMF within the mass range $m_{\rm low} = 0.01$\,M$_\odot$ to
$m_{\rm up} = 50$\,M$_\odot$. 

We note that in this paper, the choice of the mass distribution is unimportant, because we are comparing the positions of the 10 most massive stars to the positions of all stars in a region. In this case, we only require the massive stars to have masses above those of the remaining 290 objects. 

However, stochastic sampling of the IMF for low-$N$ regions can result in very different mass distributions between realisations \citep{Parker07}, and if we were to study the subsequent dynamical evolution, differences in the relative masses of individual stars could affect the measured amount of mass segregation.

We created 20 realisations of the fractal distribution (and mass
distribution). However, in the following we concentrate on just one
typical realisation of the spatial distribution, and mass distribution
of stars. The realisation is `typical' in the sense that it nicely
highlights the advantages, and disadvantages of each of the
methods. 

For each realisation we assign the stellar masses in one of three
ways.  First the stellar masses are randomly
assigned, so no mass segregation -- whatever the definition -- should
be detected (Section~\ref{mass:ran}). We then
change the positions of the ten most massive stars so that they are
either more centrally concentrated (Section~\ref{mass:hcc}) or in the
locations of the highest stellar surface density
(Section~\ref{mass:hsd}).

We perform all the subsequent analysis in 2 dimensions in order to
mimic the information available to observers.

It should be noted that {\em all} methods make a hidden assumption
that the 2 dimensional distributions are a good representation of the
true 3 dimensional structure (in the sense that they retain the same
information on spatial distributions).  It is unclear if this is
really the case, and we will return to this difficult question in a
later paper.  For now we will proceed under the assumption that the 2
dimensional distributions do retain the important information present
in the true 3 dimensional structure.

\subsection{Random distributions of stellar masses}
\label{mass:ran}

\begin{figure}
\begin{center}
\rotatebox{270}{\includegraphics[scale=0.4]{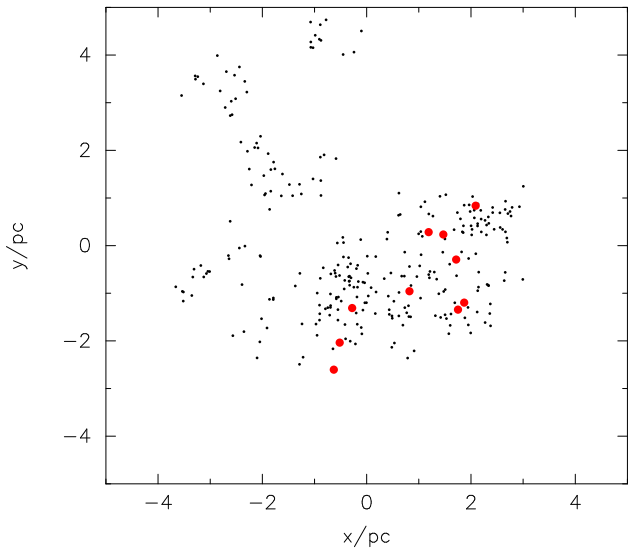}}
\end{center}
\caption[bf]{A fractal distribution ($D = 2.0$) with stars randomly
  drawn from an initial mass function and placed randomly in the
  spatial distribution. The ten most massive stars are shown by the
  larger (red) points.}
\label{map_ran}
\end{figure}

In this section we use our four techniques for measuring mass
segregation on a random distribution of stars, as shown in
Fig.~\ref{map_ran}. The locations of the ten most massive stars are
shown by the large red dots.

\begin{figure*}
  \begin{center}
\setlength{\subfigcapskip}{10pt}
\hspace*{-0.3cm} \subfigure[$\mathcal{M}_{\rm
    MF}$]{\label{mass_seg_ran-a}\rotatebox{270}{\includegraphics[scale=0.24]{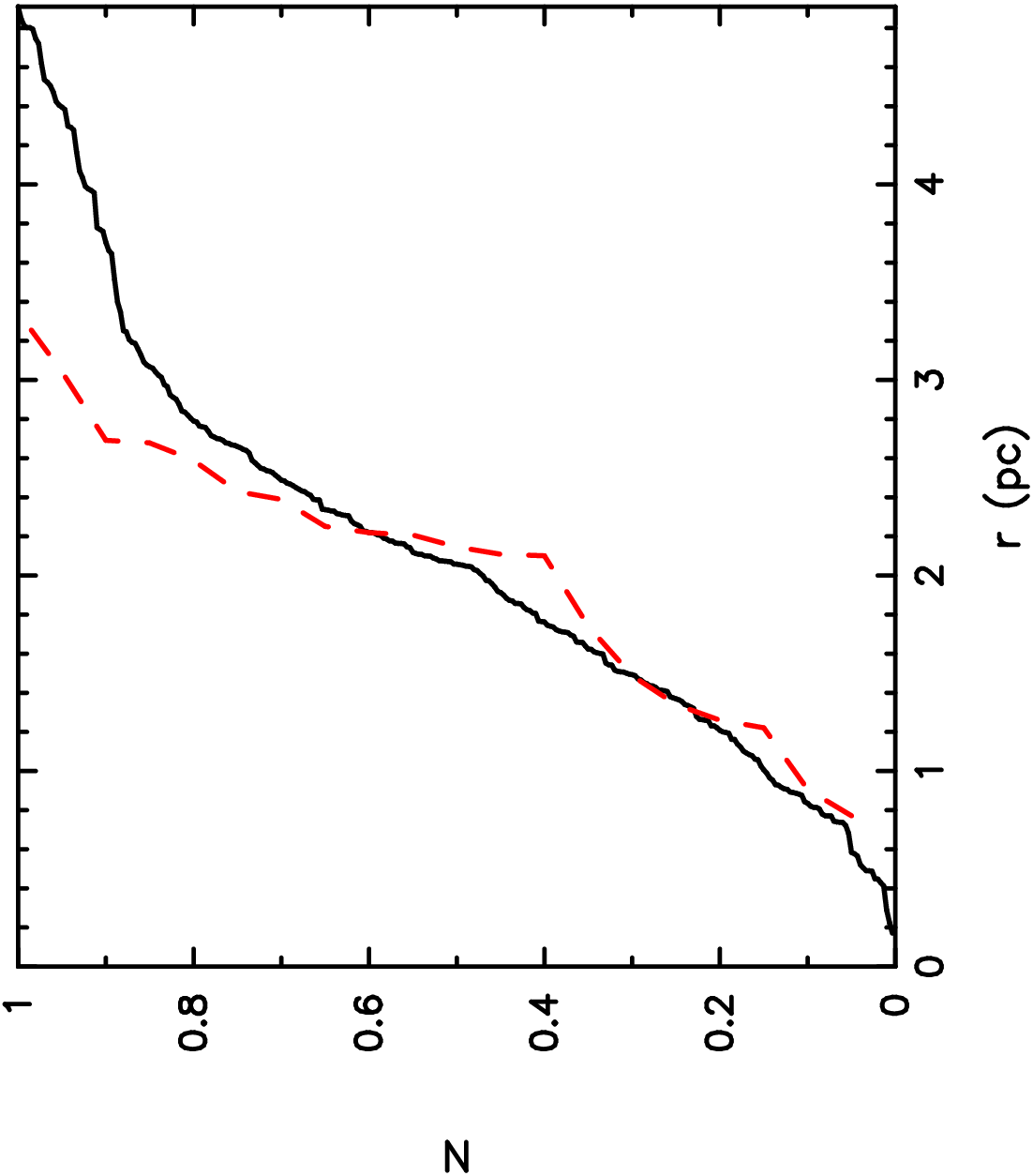}}}  
\hspace*{0.5cm} \subfigure[$\Lambda_{\rm
    MSR}$]{\label{mass_seg_ran-b}\rotatebox{270}{\includegraphics[scale=0.24]{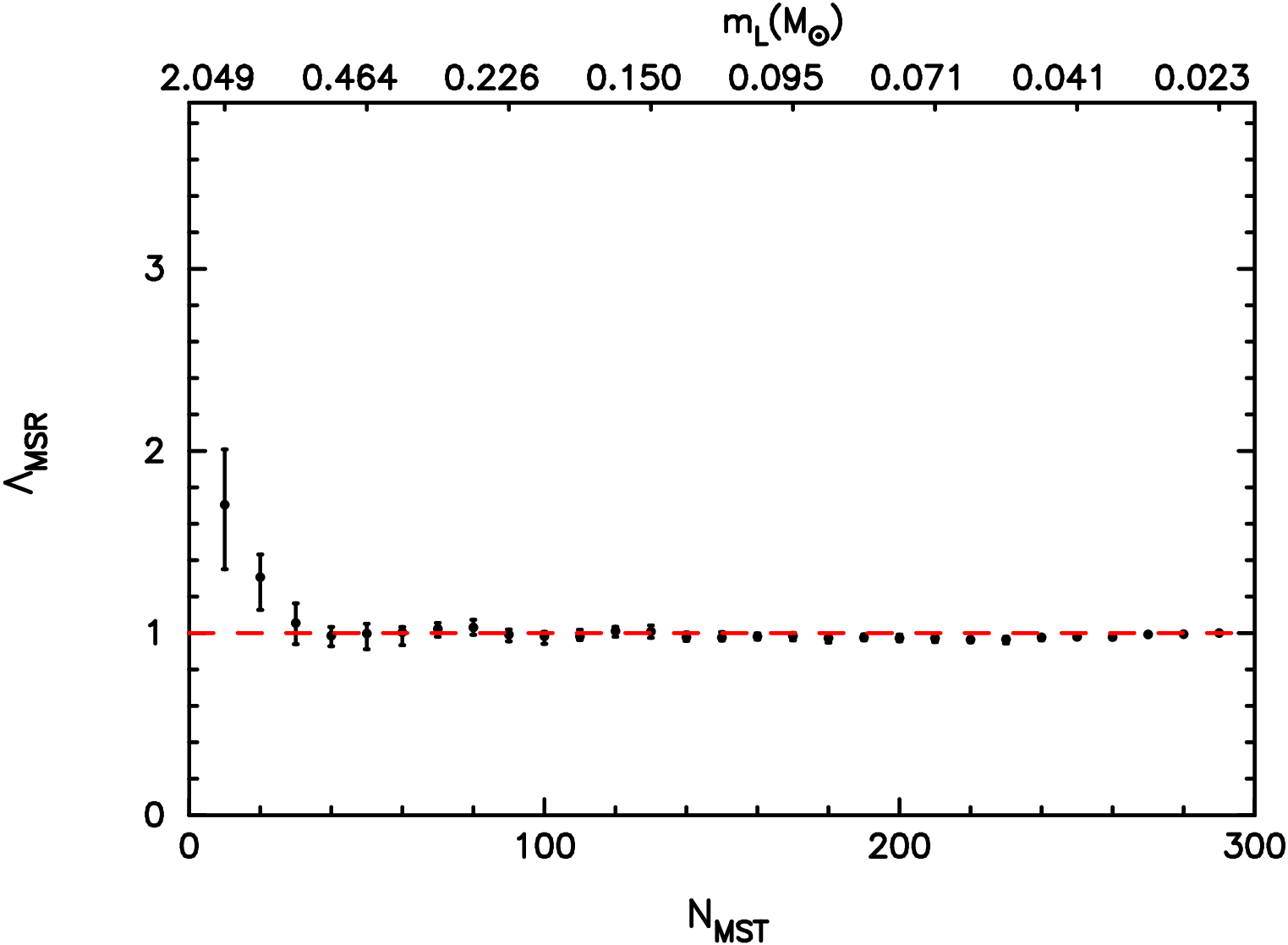}}}  
\hspace*{0.5cm} \subfigure[$\Sigma_{\rm
    LDR}$]{\label{mass_seg_ran-c}\rotatebox{270}{\includegraphics[scale=0.24]{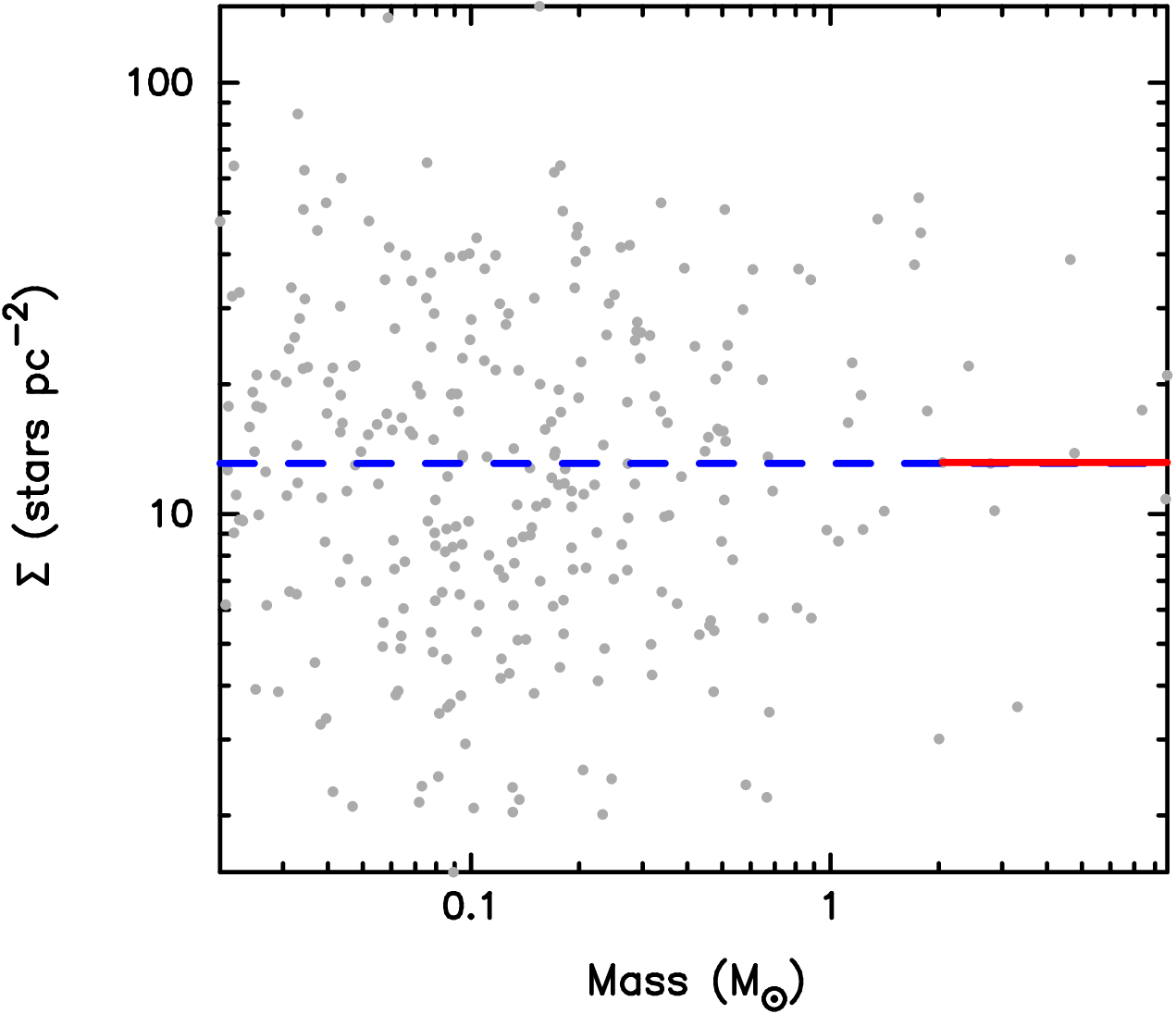}}}
\end{center}
  \caption[bf]{Three separate measures of mass segregation for the
    stellar distribution shown in Fig.~\ref{map_ran}. In panel (a) we
    show the cumulative distribution of the distance from the centre
    for the ten most massive stars (the red dashed line) and the
    cumulative distribution for all stars (the solid black line) --
    the mass function comparison, $\mathcal{M}_{\rm MF}$. In panel (b)
    we show the $\Lambda_{\rm MSR}$ mass segregation ratio as a
    function of the $N_{\rm MST}$ stars used in the subset (the lowest
    mass star, $m_L$, for various $N_{\rm MST}$ values is shown along
    the top horizontal axis). $\Lambda_{\rm MSR} = 1$ (i.e.\,\,no
    preferred spatial distribution) is shown by the solid horizontal
    red dashed line. In panel (c) we show local stellar surface
    density versus stellar mass (the $\Sigma - m$ plot). The median
    stellar surface density for the ten most massive stars is shown by
    the righthand solid (red) horizontal line and the median surface
    density for all of the stars is shown by the blue horizontal
    dashed line.}
  \label{mass_seg_ran}
\end{figure*}

\begin{figure*}
  \begin{center}
\setlength{\subfigcapskip}{10pt}
\hspace*{-0.3cm} \subfigure[MST of full
  region]{\label{kirk_omega_ran-a}\rotatebox{270}{\includegraphics[scale=0.4]{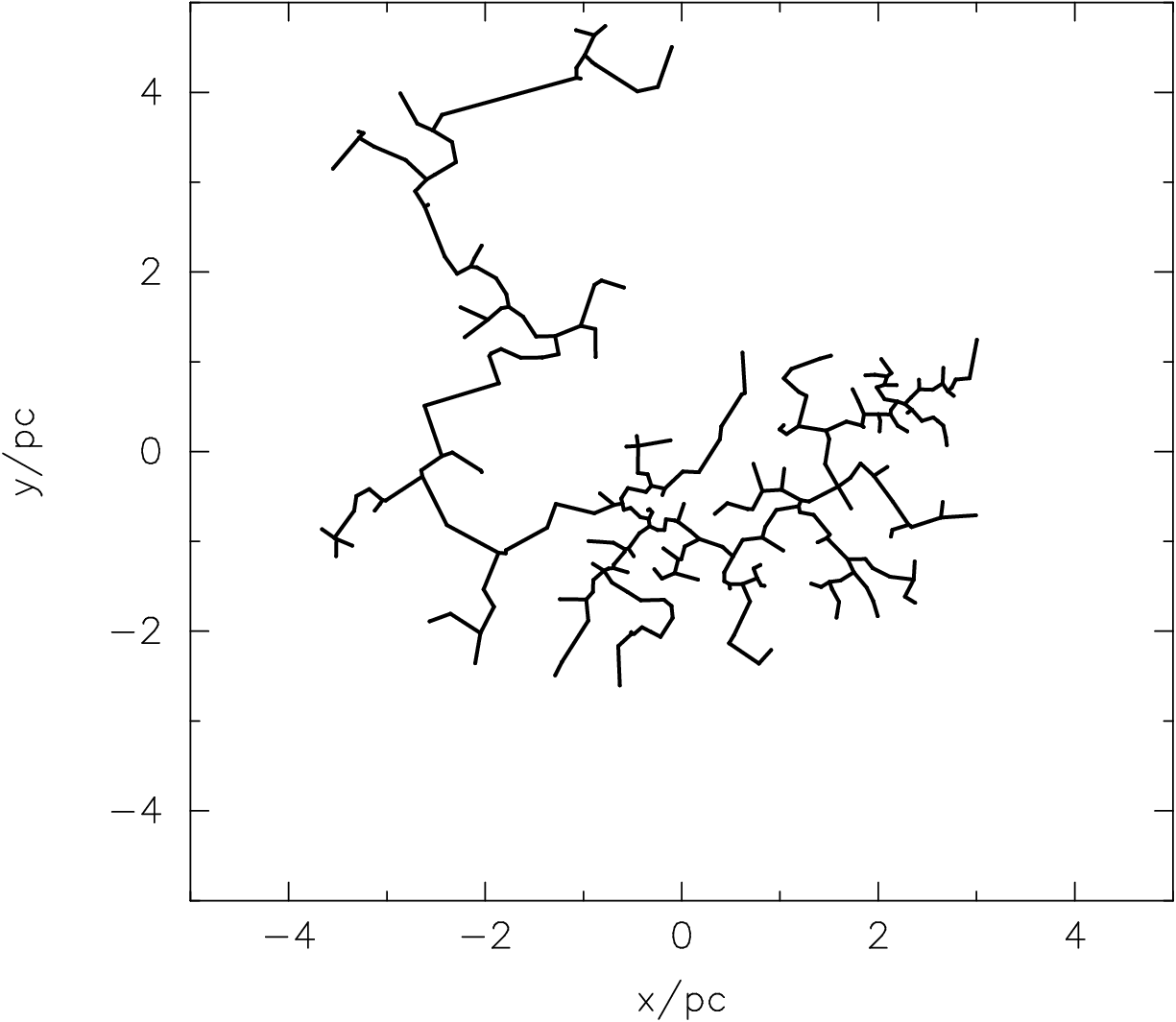}}}  
\hspace*{0.5cm} \subfigure[Distribution of branch lengths to find
  $d_{\rm
    break}$]{\label{kirk_omega_ran-b}\rotatebox{270}{\includegraphics[scale=0.4]{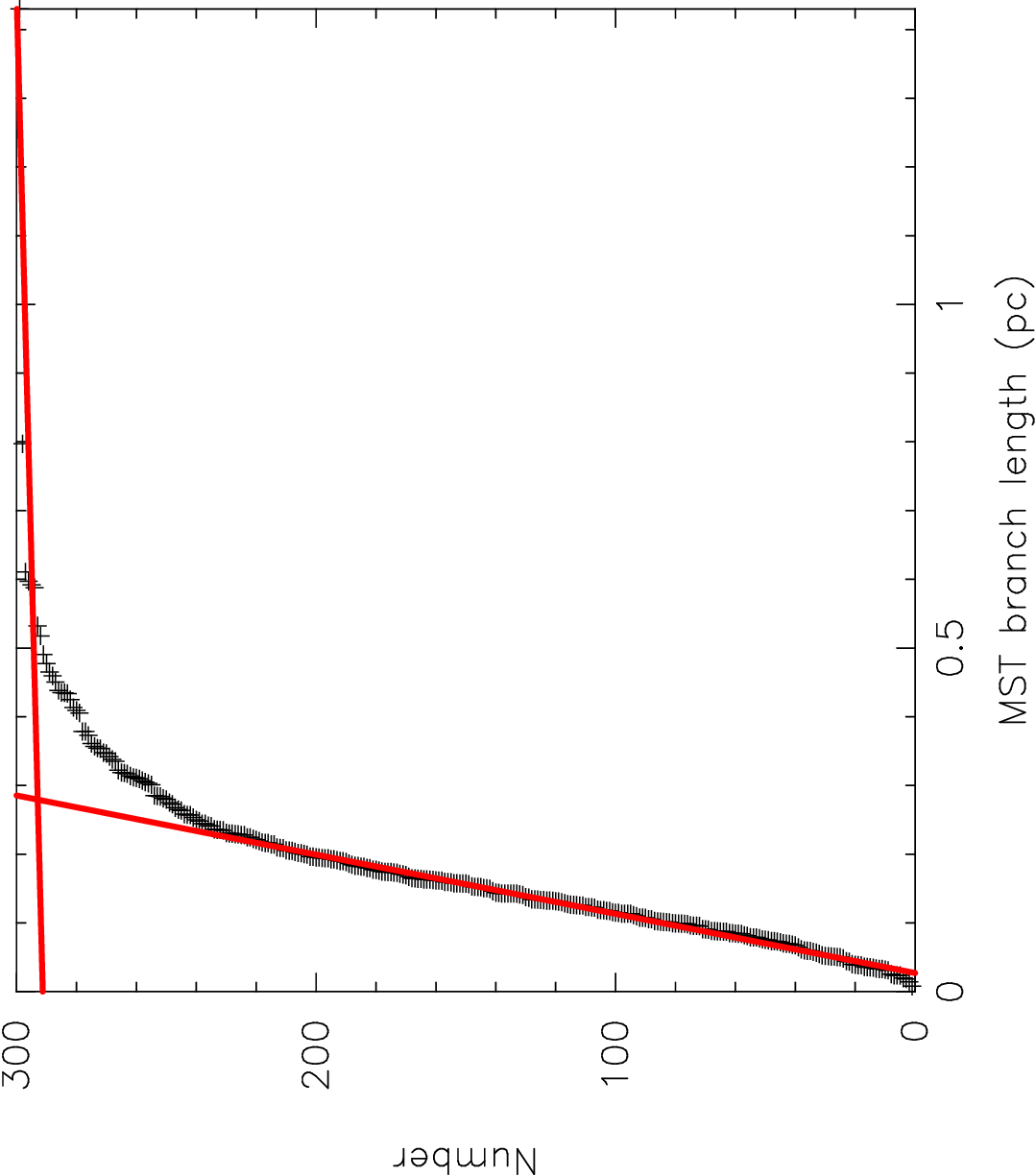}}}
\vspace*{0.25cm}
\hspace*{-0.3cm} \subfigure[Groups identified by MST $d_{\rm
    break}$]{\label{kirk_omega_ran-c}\rotatebox{270}{\includegraphics[scale=0.4]{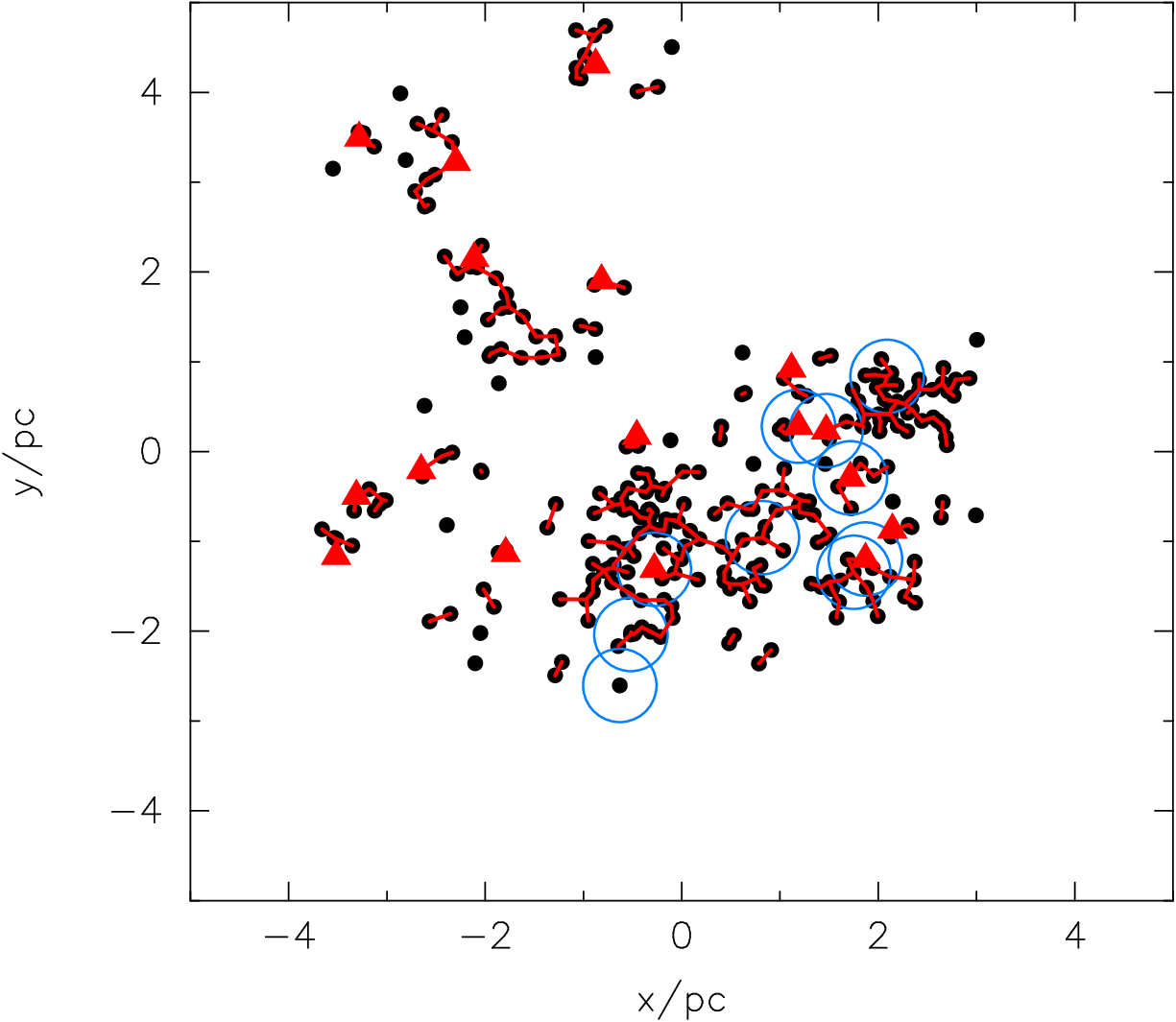}}}  
\hspace*{0.5cm} \subfigure[Mass ratio versus position offset ratio for
  groups]{\label{kirk_omega_ran-d}\rotatebox{270}{\includegraphics[scale=0.4]{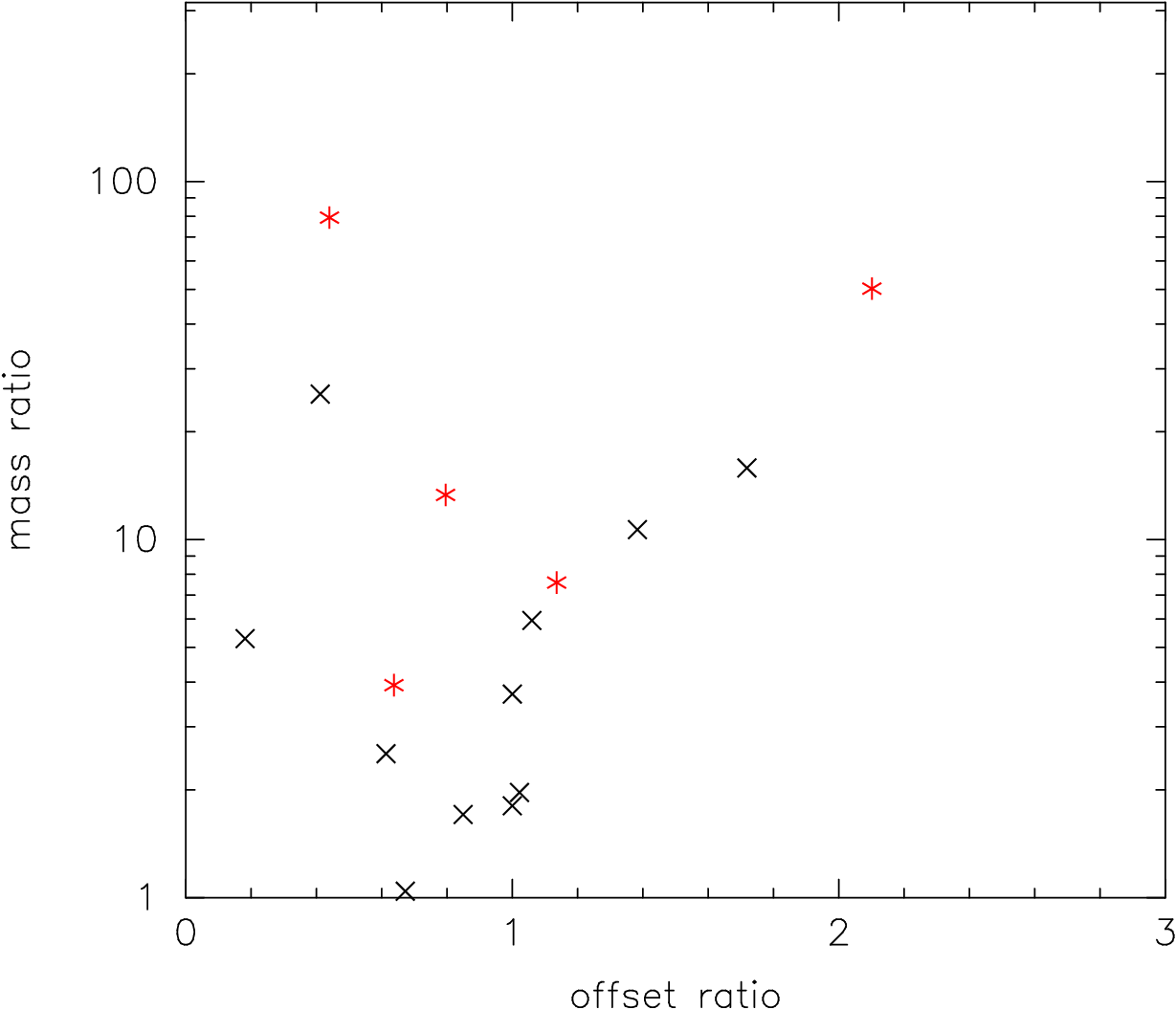}}}
\end{center}
  \caption[bf]{Mass segregation as defined by the $\Omega_{\rm GSR}$
    method. In panel (a) an MST of the full spatial distribution is
    shown, and the cumulative distribution of all of the branch
    lengths is shown in panel (b). The two power law slopes used to
    fit the data are shown by the red lines. The intersection of these
    slopes gives the critical MST length, $d_{\rm break}$, and in panel
    (c) we show the groups identified using this length. In the groups
    in which there are 3 or more stars the most massive star in the
    group is shown by the solid red triangle. The positions of the ten
    most massive stars in the \emph{full distribution} are shown by
    the large open blue circles (these correspond to the filled red
    circles in Fig.~\ref{map_ran}). In panel (d) we show the mass
    ratio of the most massive star in each group to the group median
    mass versus the ratio of the position of the most massive star to
    the median position of the group. Groups with ten or more stars
    are shown by the red asterisks.}
  \label{kirk_omega_ran}
\end{figure*} 

\subsubsection{Radial mass functions, $\mathcal{M}_{\rm MF}$}

In Fig.~\ref{mass_seg_ran-a} we show the cumulative distribution of
the radial distances from the centre of the fractal for all stars (the
solid black line) and for the ten most massive stars (the red dashed
line). Within 3\,pc of the centre, the two distributions are
overlaid. However, there are no massive stars at radii greater than
3.2\,pc and the cumulative distributions differ beyond this
radius. However, a KS test on the two distributions returns a
$p$-value of 0.51 that the two populations share the same parent
distribution, i.e.\,\,the difference is not significant in that it is higher than our threshold of $p = 0.1$.

It is important to note that the centre of the distribution is known
to us to be at $\{0,0\}$\,pc. When confronted with a distribution similar to that shown in Fig.~\ref{map_ran}, an observer would use the average position of all the stars to define a centre. In the example shown here, this average position is almost identical to the centre of mass.
 
\subsubsection{Mass segregation ratio, $\Lambda_{\rm MSR}$}
\label{mass:ran:lambda}

In Fig.~\ref{mass_seg_ran-b} we show $\Lambda_{\rm MSR}$ as a function
of the $N_{\rm MST}$ stars for the fractal region in
Fig~\ref{map_ran}. $\Lambda_{\rm MSR} = 1$ (consistent with there
being no mass segregation) is shown by the horizontal red dashed
line. The $\Lambda_{\rm MSR}$ technique shows that the 10 most massive
stars are slightly more centrally concentrated than the average stars,
with $\Lambda_{\rm MSR} = 1.7^{+0.3}_{-0.4}$ for stars with $m
>2.05$. The 20 most massive stars are also slightly more centrally
concentrated than the average stars.

This positive signal of mass segregation is likely due to the same
spatial feature in Fig.~\ref{map_ran} that shows an apparent difference in
the radial mass functions, namely that none of the most massive stars
are more than 3\,pc from the centre (Fig.~\ref{mass_seg_ran-a}).  This is a 2-$\sigma$ difference
from unity, and so would be expected roughly 1-in-20 times.  By
`fluke' this is the only random realisation that shows a 2-$\sigma$
signature of mass segregation and emphasises the need to avoid
over-interpreting a single 2-$\sigma$ result.

\subsubsection{Local density ratio, $\Sigma_{\rm LDR}$}

In Fig.~\ref{mass_seg_ran-c} we plot the local stellar surface density, $\Sigma$, against
individual stellar mass $m$ in Fig.~\ref{mass_seg_ran-c}. The median
stellar surface density for the entire distribution is
13.1\,stars\,pc$^{-2}$ (the blue dashed line) whereas the median
stellar surface density for the ten most massive stars is
13.2\,stars\,pc$^{-2}$ (the solid red line). These values are
obviously very similar, and a KS test returns a $p$-value of 0.3 that
they share the same parent distribution -- i.e. this is not a significant difference compared to our threshold value of $p = 0.1$. $\Sigma_{\rm LDR} = 1.0$ and
we therefore conclude that the most massive stars are not mass
segregated according to this definition.  

\subsubsection{Group segregation ratio, $\Omega_{\rm GSR}$}
\label{mass:ran-gsr}

In Fig.~\ref{kirk_omega_ran} we show the determination of the
$\Omega_{\rm GSR}$ group segregation ratio. We start by constructing
an MST of the entire region, as shown in
Fig.~\ref{kirk_omega_ran-a}. The cumulative distribution of the
branches in the entire MST is shown in  Fig.~\ref{kirk_omega_ran-b},
and the two power law fits to the close branches and the long branches
are shown by the solid red lines. Following \citet{Gutermuth09,Kirk10}
and \citet{Kirk14} we take the intersection of these lines as the MST
$d_{\rm break} = 0.28$~pc, the boundary between `clustered' and `diffusely'
distributed stars. All MST branches with length $< d_{\rm break}$ are
retained (Fig.~\ref{kirk_omega_ran-c}) which defines groups within the
distribution\footnote{Other methods to define groups/clusters in crowded fields may have advantages over the MST technique -- see \citet{Schmeja11} for a review.}. 

In Fig.~\ref{kirk_omega_ran-c} we show the groups (with $N>2$) selected by this
method as the stars still connected by the red MST links.  The most
massive star {\em within each of the groups} is marked by a red
triangle.  The ten most massive stars in the entire region are shown
by the large open blue circles.

It might be considered that $d_{\rm break} = 0.28$~pc has a physical
importance -- it is the apparent break between structures.  However,
in this simulation this distance has no physical significance, it is
just a projected 5~pc radius $D = 2.0$ box fractal distribution, which by design is hierarchical and self-similar, with no special spatial scale (apart from the radius itself).
Examination of Fig.~\ref{kirk_omega_ran} shows that there is nothing
`special' about this distance. Furthermore, small changes to $d_{\rm break}$ can drastically affect the number of groups which are identified. For example, if we choose $d_{\rm break} = 0.25$\,pc (the point at which the cumulative distribution in Fig.~\ref{kirk_omega_ran-b} deviates from the steep power-law), we identify 18 groups (instead of the 16 identified using $d_{\rm break} = 0.28$~pc). If we choose  $d_{\rm break} = 0.5$\,pc (where the shallow power-law slope deviates from the tail of the distribution), we identify only 6 groups.

Examination of Fig.~\ref{map_ran} shows to the eye perhaps five groups,
the most significant being to the bottom right.
Fig.~\ref{kirk_omega_ran-c} shows that $\Omega_{\rm GSR}$ has
identified many more groups than this.  In particular, the stars to the
bottom left (around $\{-3,-1\}$\,pc) have been split into three groups.  And
the significant distribution of stars at the bottom centre/right have been
split into several groups, but some stars (including one of the most
massive in the region) have not been included in any group.

The identification of groups is crucial to the $\Omega_{\rm GSR}$
method, but it is unclear from Fig.~\ref{kirk_omega_ran-c} that the
selected groups are `real' in any sense.

There are three other significant issues with the $\Omega_{\rm GSR}$ method
that are immediately apparent from Fig.~\ref{kirk_omega_ran-c}.

Firstly, the most massive star in a group may not be one of the most
massive stars in the region.  All of the groups to the upper left have
a locally most massive star that is not one of the most massive stars
in the region as a whole.

Secondly, a group may contain more than one of the truly most massive
stars in a region (e.g. three of the larger groups to the bottom
right) in which case only the most massive of these is considered and
the other (truly massive for the region) stars are discarded.

Thirdly, if a truly massive star is not part of a group (surely an
interesting phenonena) then it is discarded from the analysis entirely
(e.g. the large open blue circle at the bottom centre).

But taking the $\Omega_{\rm GSR}$ method to its conclusion we show the 
mass ratio (most massive star in the group to median star
in the group) versus offset ratio (position of most massive star to
median group position) in for all groups with $N > 2$ stars in
Fig.~\ref{kirk_omega_ran-d}. The five groups with $N \geq 10$ stars
are shown by the red points, and three of them have an offset ratio
less than unity, i.e.\,\,they are mass segregated according to the
definition in \citet{Kirk10} and \citet{Kirk14}. We define the group
segregation ratio as the number of groups with an offset ratio less
than or equal to unity divided by the total number of groups. For all
groups with    $N > 2$, $\Omega_{\rm GSR} = 0.59$ and for groups with
$N \geq 10$,   $\Omega_{\rm GSR} = 0.60$\footnote{Note that \citet{Kirk10} and \citet{Kirk14} generally only present statistics for groups containing 10 or more stars, and we will also draw conclusions based only on these `large' groups.}. In a truly random
distribution, $\Omega_{\rm GSR} = 0.5$.

It is worth noting another problem here, in that the $\Omega_{\rm
  GSR}$ method needs to define a `centre' of each group from which to
measure distances.  Therefore there is an implicit assumption of
spherical symmetry which examination of Fig.~\ref{kirk_omega_ran-c}
shows not to be the case in most groups.

There are two groups with $N \geq 40$ stars, which is enough points to
run the $\Lambda_{\rm MSR}$ and $\Sigma_{\rm LDR}$ methods on these
groups in isolation. One of these groups has an offset ratio of less
than unity (i.e.\,\,it is mass segregated according to this method) --
however, neither  $\Lambda_{\rm MSR}$ nor $\Sigma_{\rm LDR}$ find mass
segregation in this group. The second group has  an offset ratio of
greater than unity (i.e.\,\,it is \emph{inversely} mass segregated
according to this method). Again, both  $\Lambda_{\rm MSR}$ and
$\Sigma_{\rm LDR}$ are consistent with no mass segregation (normal or
inverse).

\begin{figure}
\begin{center}
\rotatebox{270}{\includegraphics[scale=0.4]{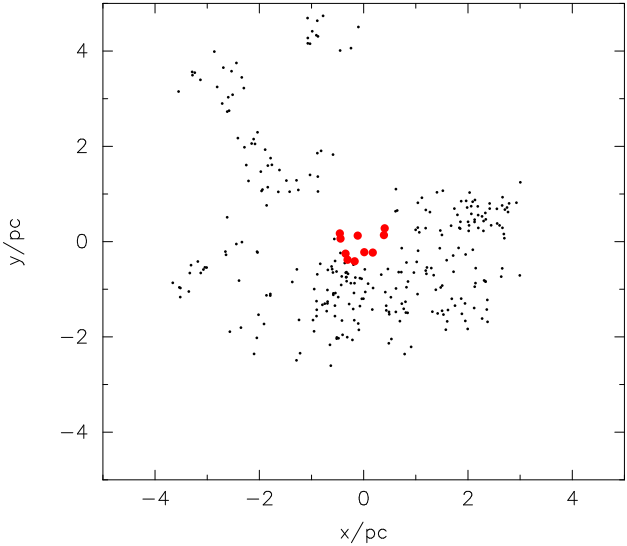}}
\end{center}
\caption[bf]{As Fig.~\ref{map_ran}; a fractal distribution of stellar
  masses randomly drawn from an initial mass function. However, in
  this case we have swapped the locations of the ten most massive
  stars (shown by the larger red points) with the ten most central
  stars.}
\label{map_hcc}
\end{figure}

%\newpage

\begin{figure*}
  \begin{center}
\setlength{\subfigcapskip}{10pt}
\hspace*{-0.3cm} \subfigure[$\mathcal{M}_{\rm
    MF}$]{\label{mass_seg_hcc-a}\rotatebox{270}{\includegraphics[scale=0.24]{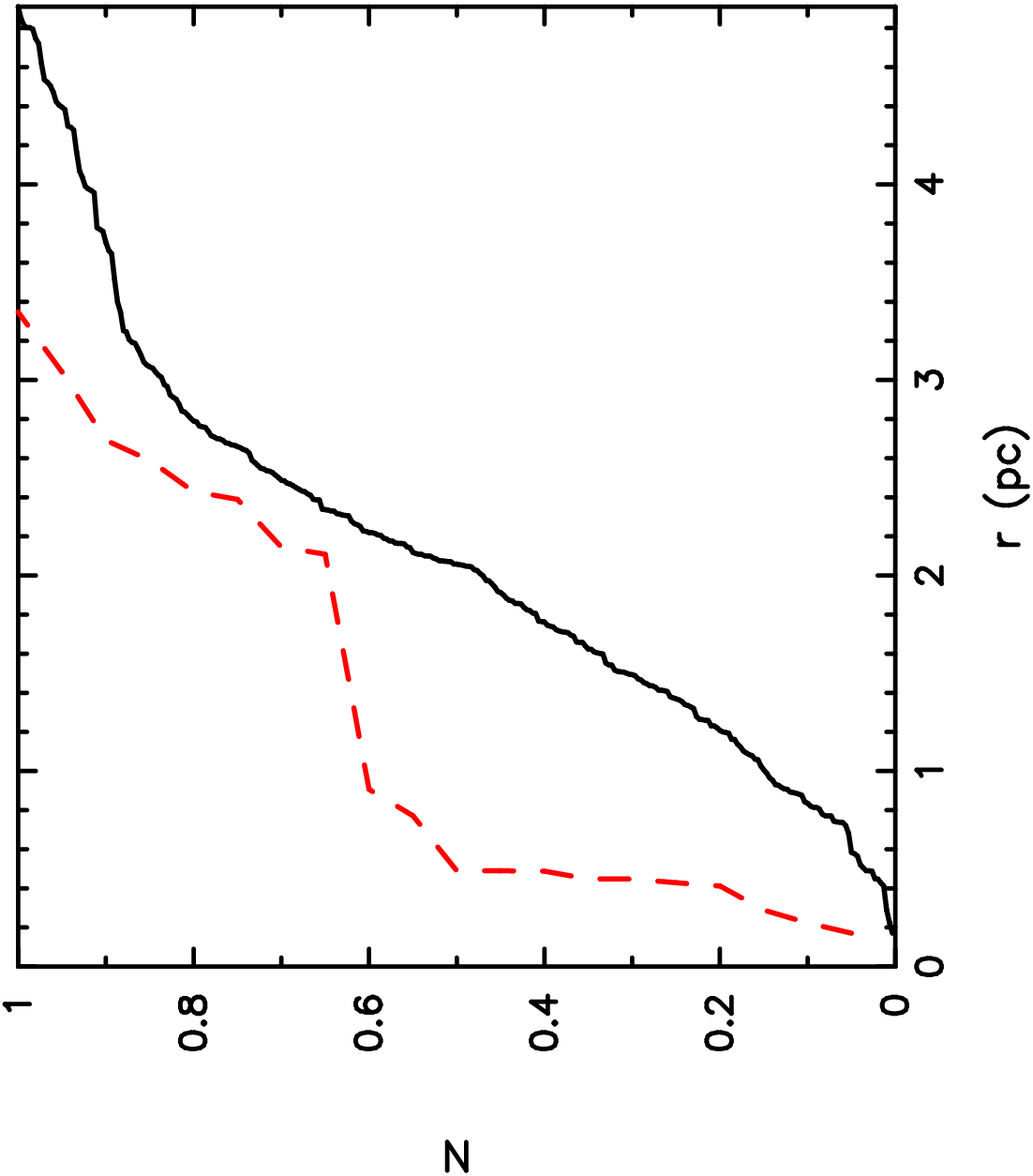}}}  
\hspace*{0.5cm} \subfigure[$\Lambda_{\rm
    MSR}$]{\label{mass_seg_hcc-b}\rotatebox{270}{\includegraphics[scale=0.24]{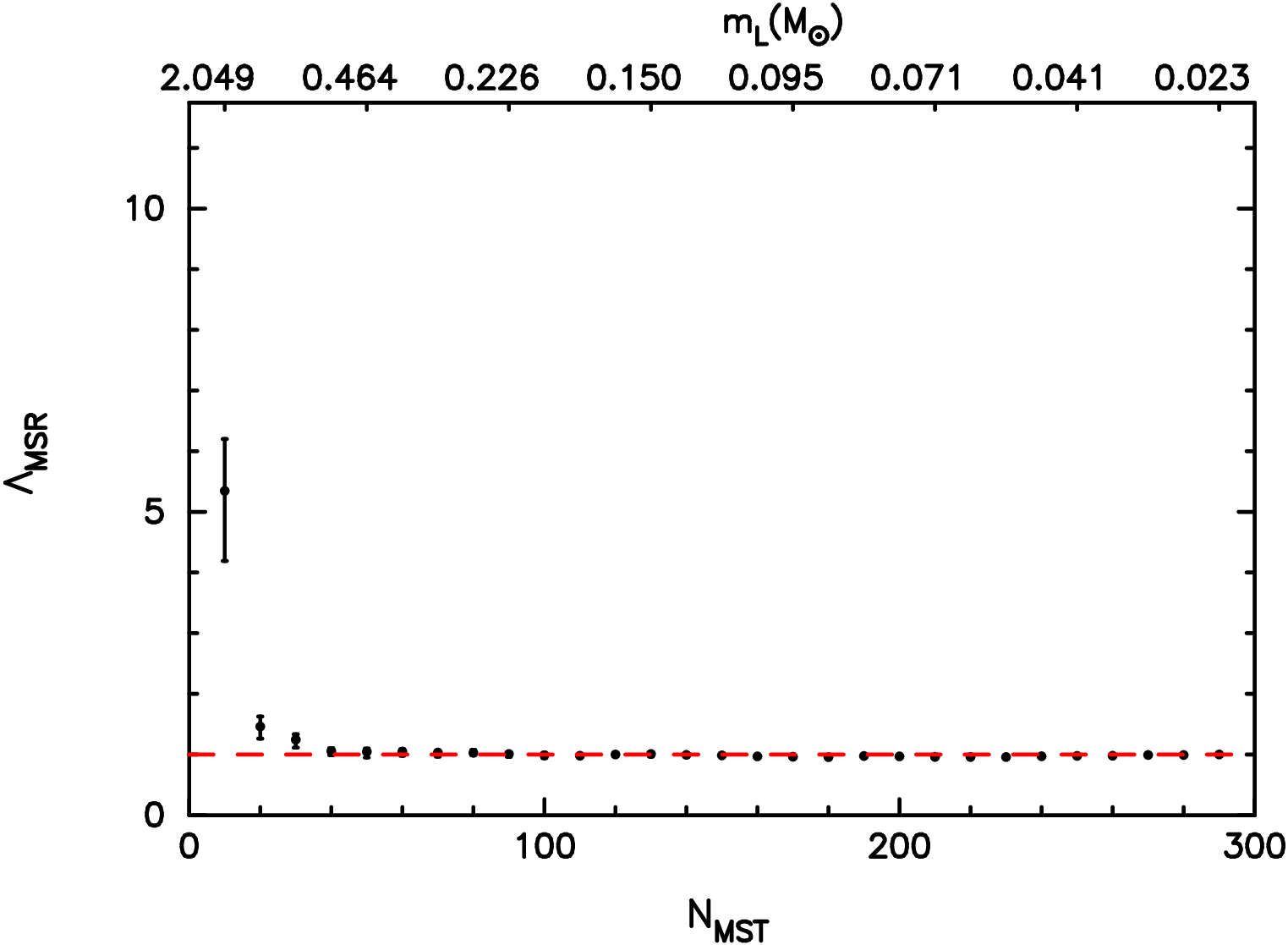}}}  
\hspace*{0.5cm} \subfigure[$\Sigma_{\rm
    LDR}$]{\label{mass_seg_hcc-c}\rotatebox{270}{\includegraphics[scale=0.24]{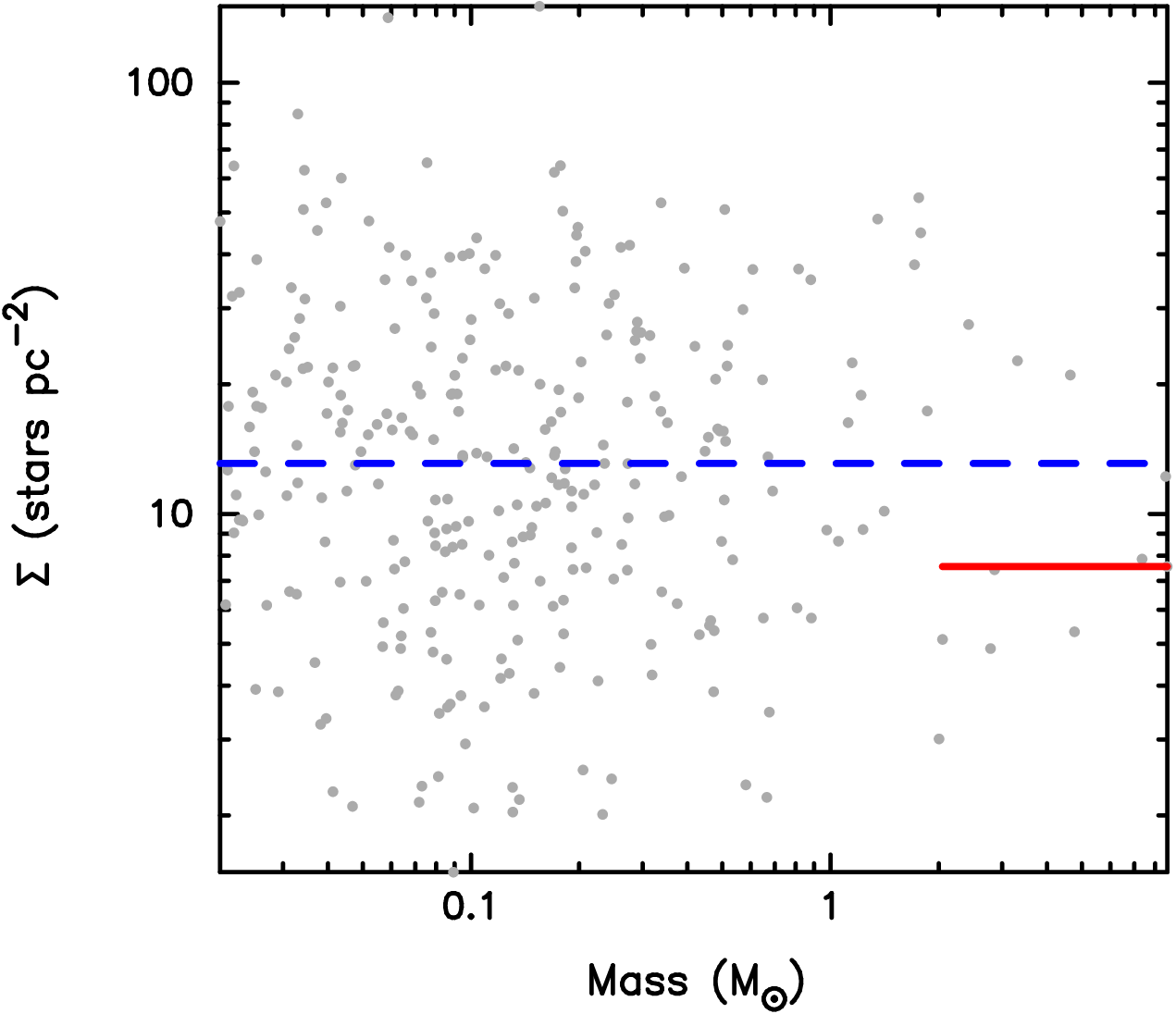}}}
\end{center}
  \caption[bf]{Three separate measures of mass segregation for the
    stellar distribution shown in Fig.~\ref{map_hcc} where the most
    massive stars are centrally concentrated. In panel (a) we show the
    cumulative distribution of the distance from the centre for the
    ten most massive stars (the red dashed line) and the cumulative
    distribution for all stars (the solid black line) -- the mass
    function comparison, $\mathcal{M}_{\rm MF}$. In panel (b) we show
    the $\Lambda_{\rm MSR}$ mass segregation ratio as a function of
    the $N_{\rm MST}$ stars used in the subset (the lowest mass star,
    $m_L$, for various $N_{\rm MST}$ values is shown along the top
    horizontal axis). $\Lambda_{\rm MSR} = 1$ (i.e.\,\,no preferred
    spatial distribution) is shown by the solid horizontal red dashed
    line. In panel (c) we show local stellar surface density versus
    stellar mass (the $\Sigma - m$ plot). The median stellar surface
    density for the ten most massive stars is shown by the righthand
    solid (red) horizontal line and the median surface density for all
    of the stars is shown by the blue horizontal dashed line.}
  \label{mass_seg_hcc}
\end{figure*}

%\newpage

\begin{figure*}
  \begin{center}
\setlength{\subfigcapskip}{10pt}
\hspace*{-0.3cm} \subfigure[Groups identified by MST $d_{\rm
    break}$]{\label{kirk_omega_hcc-c}\rotatebox{270}{\includegraphics[scale=0.4]{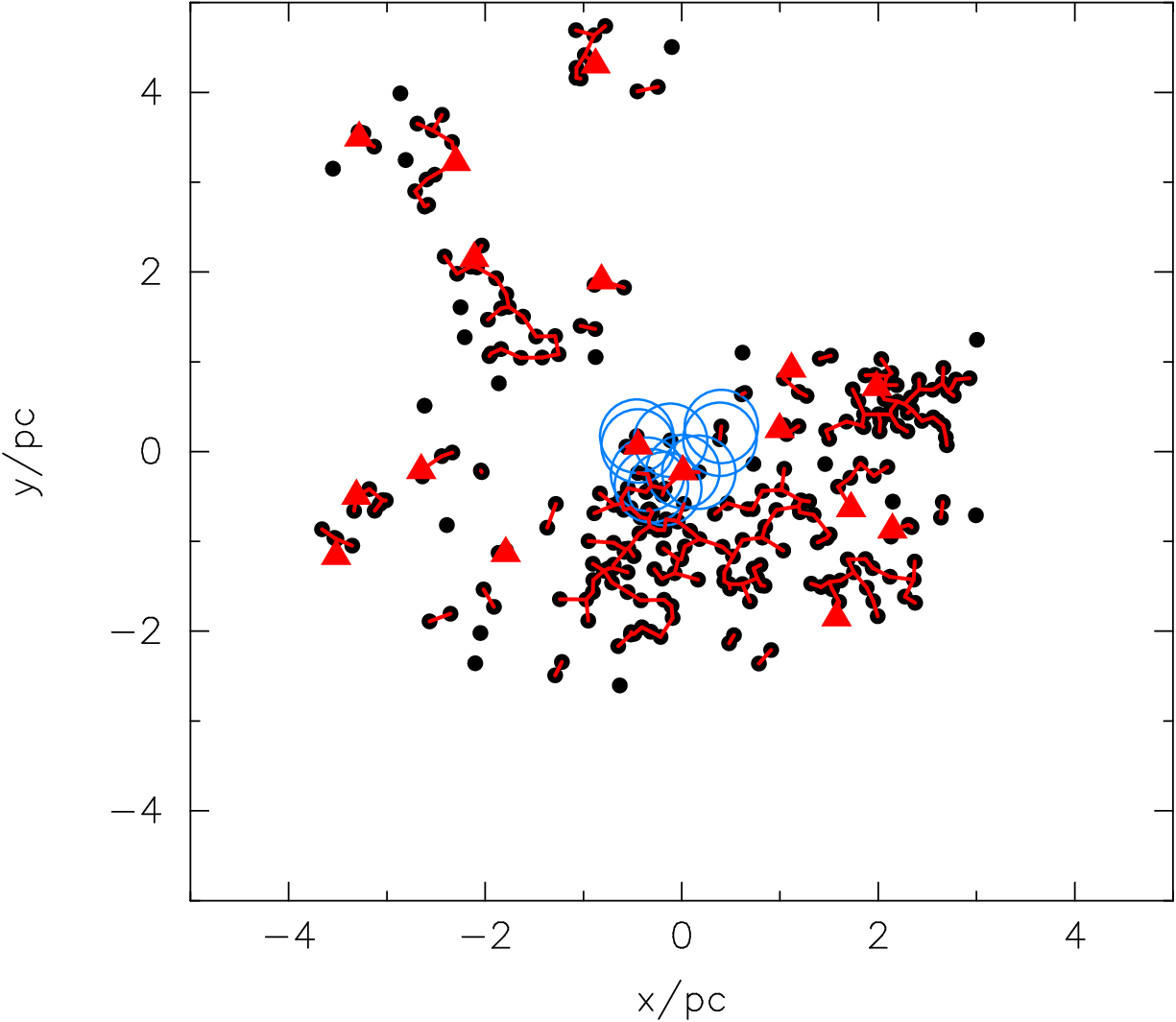}}}  
\hspace*{0.5cm} \subfigure[Mass ratio versus position offset ratio for
  groups]{\label{kirk_omega_hcc-d}\rotatebox{270}{\includegraphics[scale=0.4]{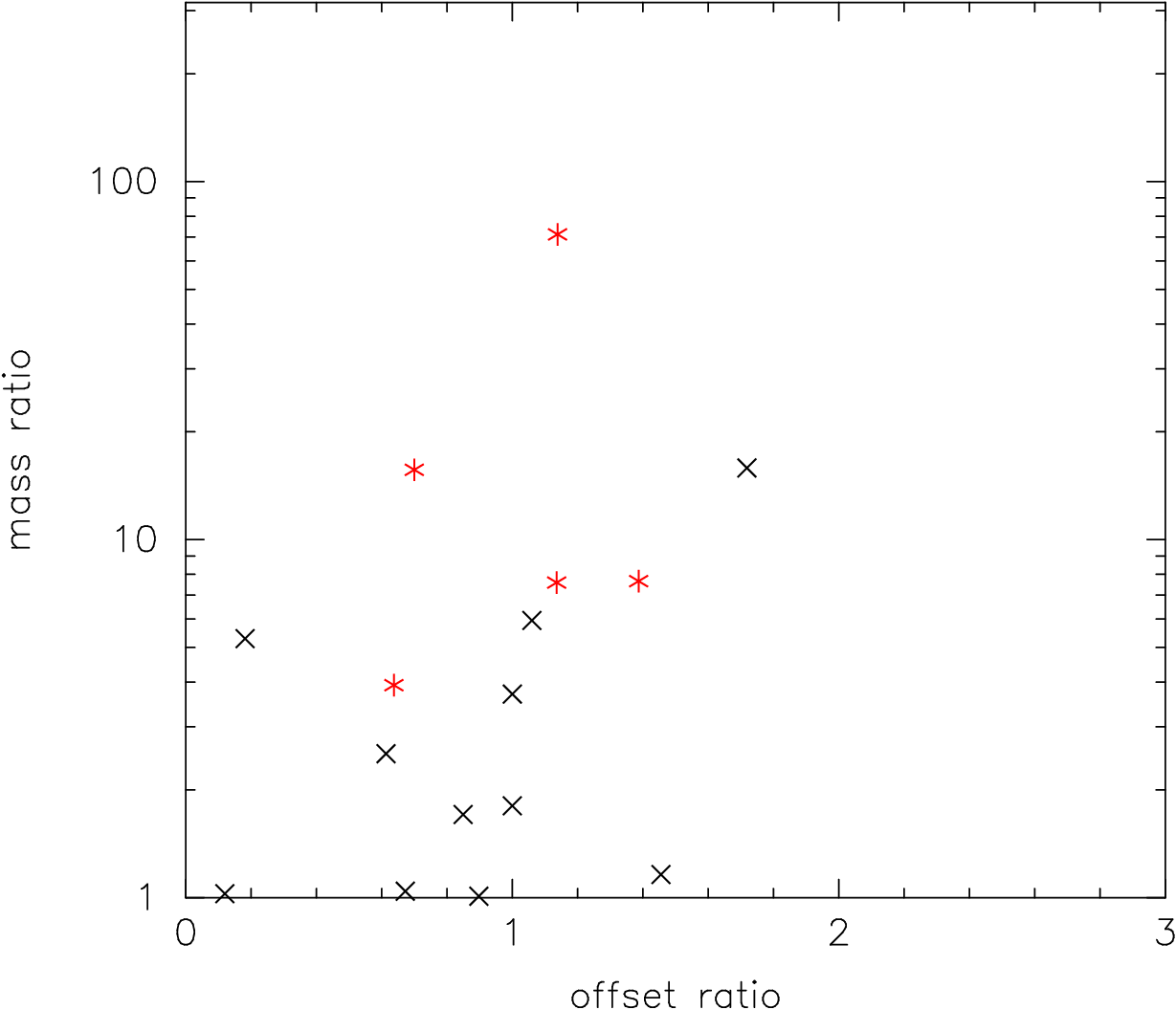}}}
\end{center}
  \caption[bf]{Mass segregation as defined by the $\Omega_{\rm GSR}$
    method for the stellar distribution shown in Fig.~\ref{map_hcc}
    where the most massive stars are centrally concentrated. The
    stellar groups are identified as shown in
    Fig.~\ref{kirk_omega_ran-a} and Fig.~\ref{kirk_omega_ran-b}. In
    panel (a) we show the groups identified using MST $d_{\rm
      break}$. In the groups in which there are 3 or more stars the
    most massive star in the group is shown by the solid red
    triangle. The positions of the ten most massive stars in the
    \emph{full distribution} are shown by the large open blue circles
    (these correspond to the filled red circles in
    Fig.~\ref{map_hcc}). In panel (d) we show the mass ratio of the
    most massive star in each group to the group median mass versus
    the ratio of the position of the most massive star to the median
    position of the group. Groups with ten or more stars are shown by
    the red asterisks.}
  \label{kirk_omega_hcc}
\end{figure*}

\subsection{Massive stars centrally concentrated}
\label{mass:hcc}

We now swap the positions of the 10 most massive stars with the
positions of the 10 stars closest to the centre of the fractal
distribution, as shown by the red points in Fig.~\ref{map_hcc}.  This
is clearly a rather artificial distribution of the most massive stars,
but it is one that most closely matches the `classical' definition of
mass segregation for this region.

\subsubsection{Radial mass functions, $\mathcal{M}_{\rm MF}$}

In Fig.~\ref{mass_seg_hcc-a} the cumulative distribution of radial
positions of the 10 most massive stars is shown by the red dashed
line, whereas the cumulative distribution of the radial positions for
all stars is shown by the solid black line. Due to the central
concentration of the most massive stars, the KS test returns a
$p$-value of $2 \times 10^{-4}$ that the two subsets share the same
parent distribution.

This result demonstrates that if we have confidence in the definition of the centre of a region, a strong mass segregation signature may still be seen in substructured distributions using the radial mass function technique.

\subsubsection{Mass segregation ratio, $\Lambda_{\rm MSR}$}

We show the  $\Lambda_{\rm MSR}$ ratio in
Fig.~\ref{mass_seg_hcc-b}. The ten most massive stars have
$\Lambda_{\rm MSR} = 5.3^{+0.9}_{-1.0}$, which is significantly above
unity. In 20 realisations, only 1 star-forming region displays a
$\Lambda_{\rm MSR}$ ratio that is not significantly above unity, with
values ranging from $\Lambda_{\rm MSR} = 2.0^{+0.3}_{-0.4}$ to
$\Lambda_{\rm MSR} = 11.1^{+1.1}_{-1.4}$.

This is completely unsurprising as $\Lambda_{\rm MSR}$ is designed to
measure exactly this type of mass segregation -- the most massive
stars being much closer to one-another than a random sample of stars
would be.

\subsubsection{Local density ratio, $\Sigma_{\rm LDR}$}

Interestingly, the  $\Sigma_{\rm LDR}$ ratio does not reflect the
central concentration of the 10 most massive stars.  $\Sigma_{\rm LDR}
= 0.58$ due to the most centrally located stars being in areas of
relatively low surface density, although a KS test returns a $p$-value
of 0.26 that the massive stars have a different parent distribution to
the full distribution (i.e.\,\,this difference is not
significant). That said, most people would conclude simply from eye
that the distribution shown in Fig.~\ref{map_hcc} is mass segregated,
even though the massive stars have low local surface density.  
In 20 realisations of this distribution,  $\Sigma_{\rm
  LDR}$ does not detect mass segregation in 11, and in a further 5 it
finds inverse mass segregation.

\subsubsection{Group segregation ratio, $\Omega_{\rm GSR}$}

The overall spatial distribution has not changed between
Figs.~\ref{map_ran}~and~\ref{map_hcc} and so the determination of
$d_{\rm break}$ and the subsequent identification of groups is
identical to that in Section~\ref{mass:ran}. We show the groups
defined by $d_{\rm break}$ in Fig.~\ref{kirk_omega_hcc-c}, noting the
change of location of the 10 most massive stars in the full
distribution (the blue open circles). The locations of the most
massive star in each group (shown by the red triangles) have also
changed in some cases. 

Again, we note that two of the 10 most massive stars from the full
distribution are now no longer part of a group with $N>2$, and instead
are in a pair (located at $\{0.4,0.1\}$\,pc). We show the mass ratio
versus offset ratio  for all groups with $N > 2$ stars in
Fig.~\ref{kirk_omega_hcc-d}. Again, the five groups with $N \geq 10$
stars are shown by the red points. This time only two of these  five
groups are mass segregated ($\Omega_{\rm GSR} = 0.40$), even though
the global distribution of massive stars is very mass segregated. 

In this situation -- where the most massive stars in a region are
`centrally' concentrated, it is not clear what $\Omega_{\rm GSR}$ is
measuring.  The majority of the measurements of mass ratio versus
offset ratio do not include any of the ten most massive stars in the
region.  

\subsection{Massive stars in areas of high density}
\label{mass:hsd}

\begin{figure}
\begin{center}
\rotatebox{270}{\includegraphics[scale=0.4]{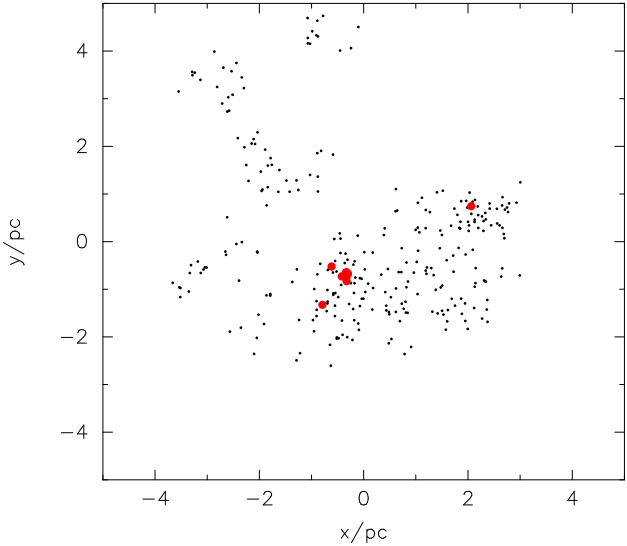}}
\end{center}
\caption[bf]{As Fig.~\ref{map_ran}; a fractal distribution of stellar
  masses randomly drawn from an initial mass function. However, in
  this case we have swapped the locations of the ten most massive
  stars (shown by the larger red points) with the ten stars with the
  highest local stellar surface densities (as defined by
  Eq.~\ref{sigma}).}
\label{map_hsd}
\end{figure}

We now change the positions of the massive stars once more, and swap
them with stars with the highest local surface densities, as shown by
the red points in Fig.~\ref{map_hsd}.  In this case we are shifting
from a type of mass segregation related to `classical' mass
segregation to one in which the massive stars are associated with the
highest density regions (at least surface density; the volume
densities of these locales would be unknown to the hypothetical
observer).  

\subsubsection{Radial mass functions, $\mathcal{M}_{\rm MF}$}

The radial mass function for the 10 most massive stars (the red dashed
line) and the whole distribution (the black solid line) is shown in
Fig.~\ref{mass_seg_hsd-a}. As for the centrally concentrated cluster,
the most massive stars are closer to the centre (none are outside of
3.3\,pc) than the average stars, and a KS test returns a $p$-value of
$7 \times 10^{-3}$ that they share the same underlying parent
distribution. However, in 13 of our 20 realisations in which we draw different masses and positions for the stars each time, the KS test
returns a $p$-value in excess of 0.1, suggesting that the massive
stars are not closer to the centre than the average members. This is not entirely surprising, as the positions in the region with the highest surface density may not be co-located, as is the case for one of the massive stars in Fig.~\ref{map_hsd}.

\subsubsection{Mass segregation ratio, $\Lambda_{\rm MSR}$}

We show the measurement of $\Lambda_{\rm MSR}$ for this distribution
in Fig.~\ref{mass_seg_hsd-b}. The 10 most massive stars have
$\Lambda_{\rm MSR} = 2.7^{+0.4}_{-0.6}$, i.e.\,\,mass segregation is
present according to this measure, but is not as strong as in the case
where we placed the most massive stars at the centre of the
distribution. 

In many ways this is not surprising.  Stars with the highest surface
density are reasonably likely to be close to one-another (as the
surface density is high), and so this artifical set-up will often
place several massive stars close to one-another (here at
$\{-0.5,-0.7\}$\,pc). This is not always the case, from 20
realisations of this distribution, in 11 the measured $\Lambda_{\rm
  MSR}$ was greater than unity, but less than two (and only
marginally significant, error bars typically being around $\pm 0.5$).

\subsubsection{Local density ratio, $\Sigma_{\rm LDR}$}

\begin{figure*}
  \begin{center}
\setlength{\subfigcapskip}{10pt}
\hspace*{-0.3cm} \subfigure[$\mathcal{M}_{\rm
    MF}$]{\label{mass_seg_hsd-a}\rotatebox{270}{\includegraphics[scale=0.24]{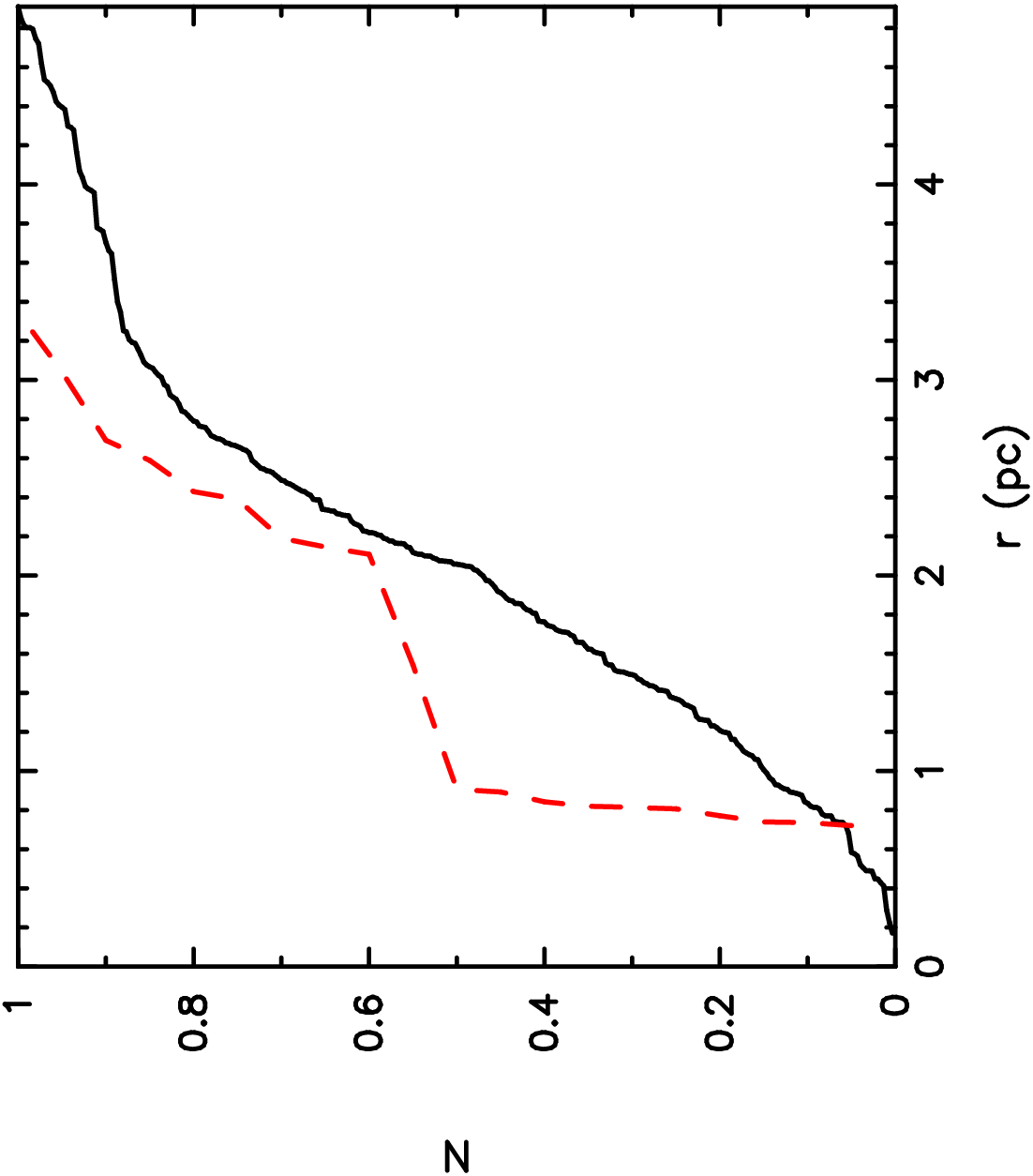}}}  
\hspace*{0.5cm} \subfigure[$\Lambda_{\rm
    MSR}$]{\label{mass_seg_hsd-b}\rotatebox{270}{\includegraphics[scale=0.24]{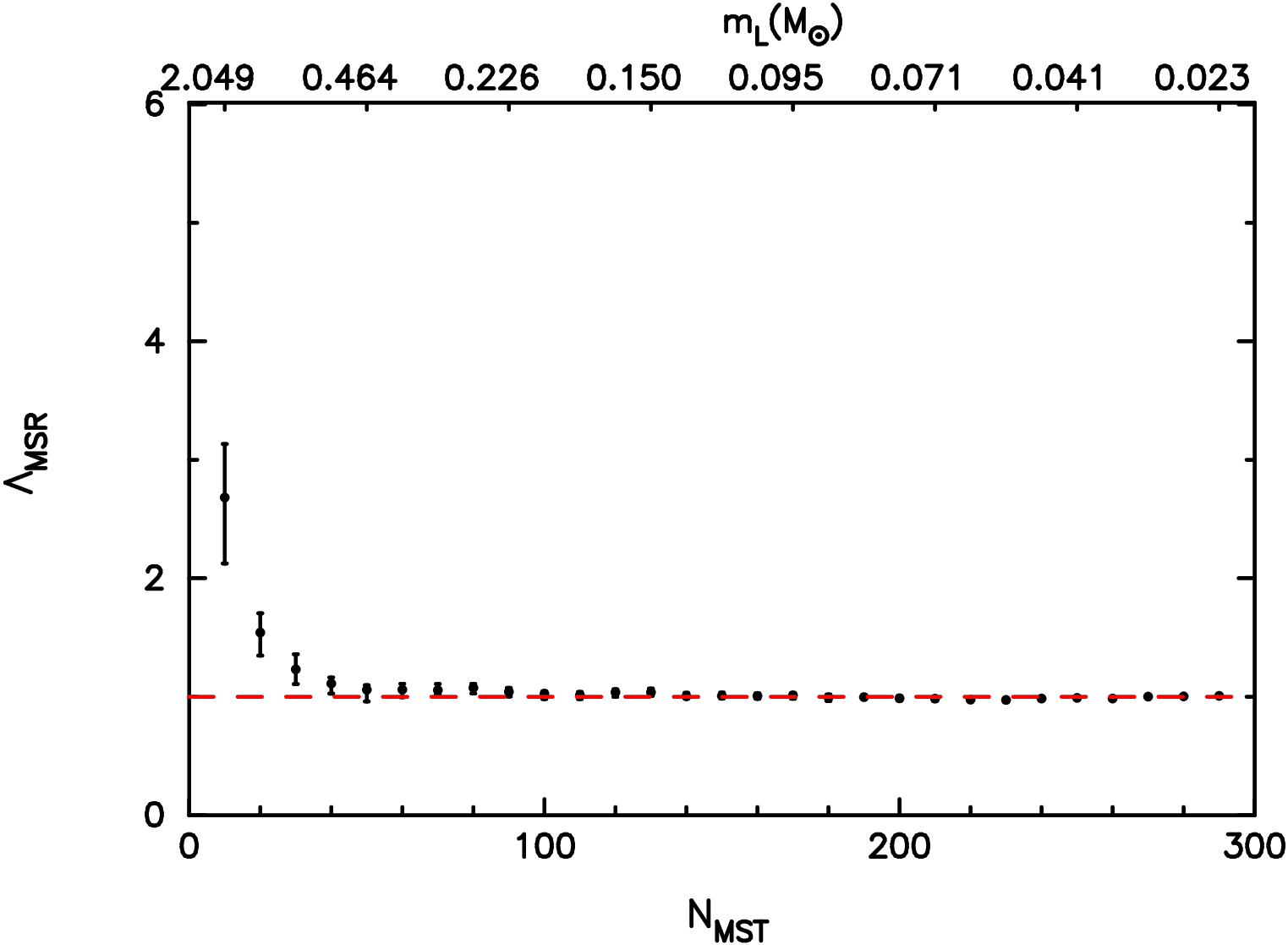}}}  
\hspace*{0.5cm} \subfigure[$\Sigma_{\rm
    LDR}$]{\label{mass_seg_hsd-c}\rotatebox{270}{\includegraphics[scale=0.24]{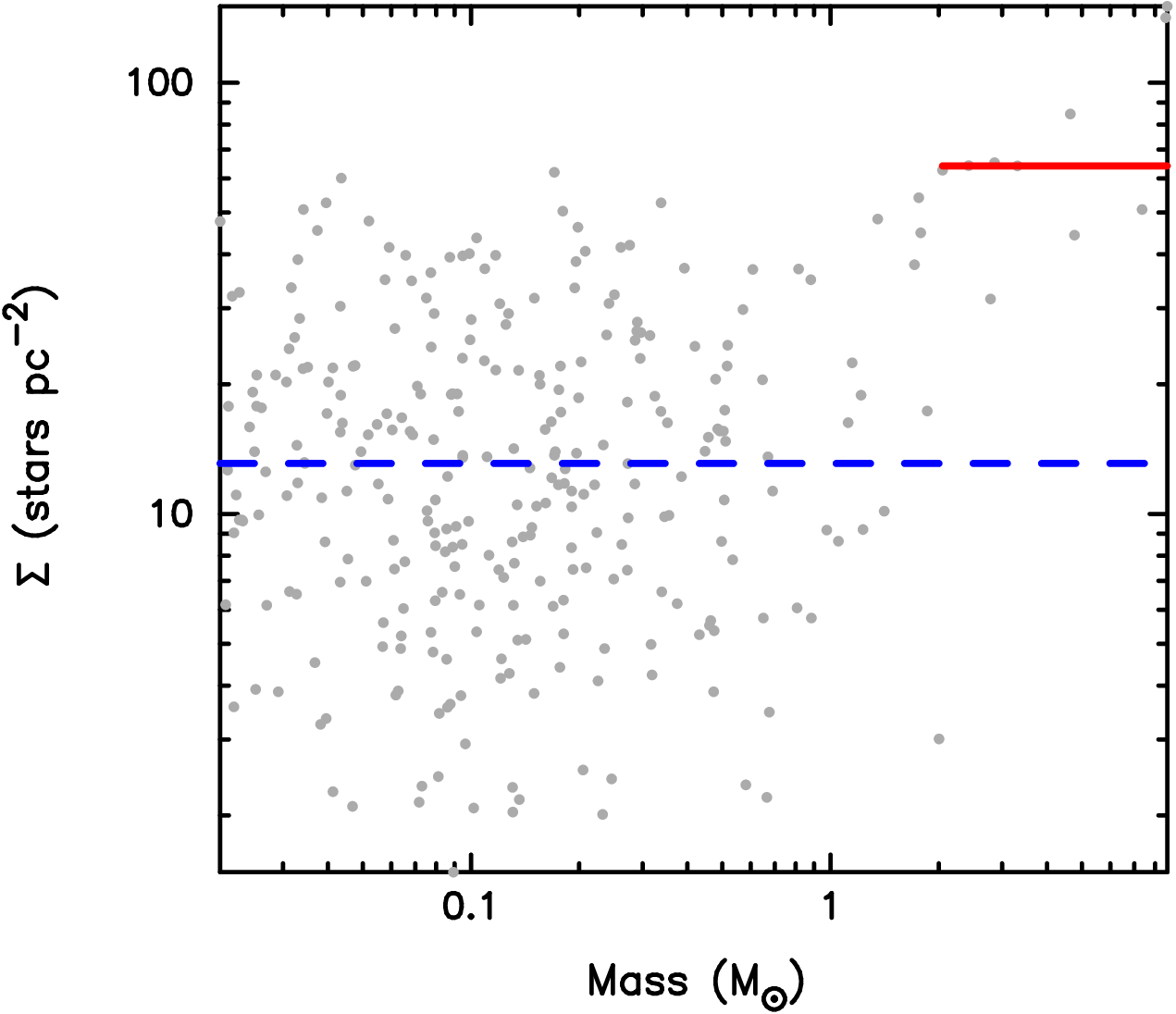}}}
\end{center}
  \caption[bf]{Three separate measures of mass segregation for the
    stellar distribution shown in Fig.~\ref{map_hsd} where the most
    massive stars are in the areas of highest stellar surface
    density. In panel (a) we show the cumulative distribution of the
    distance from the centre for the ten most massive stars (the red
    dashed line) and the cumulative distribution for all stars (the
    solid black line) -- the mass function comparison,
    $\mathcal{M}_{\rm MF}$. In panel (b) we show the $\Lambda_{\rm
      MSR}$ mass segregation ratio as a function of the $N_{\rm MST}$
    stars used in the subset (the lowest mass star, $m_L$, for various
    $N_{\rm MST}$ values is shown along the top horizontal
    axis). $\Lambda_{\rm MSR} = 1$ (i.e.\,\,no preferred spatial
    distribution) is shown by the solid horizontal red dashed line. In
    panel (c) we show local stellar surface density versus stellar
    mass (the $\Sigma - m$ plot). The median stellar surface density
    for the ten most massive stars is shown by the righthand solid
    (red) horizontal line and the median surface density for all of
    the stars is shown by the blue horizontal dashed line. }
  \label{mass_seg_hsd}
\end{figure*}

When we compare the surface density of the most massive stars to the
average surface density, unsurprisingly the most massive stars have
much higher median values, as shown in
Fig.~\ref{mass_seg_hsd-c}. Here, the most massive stars have $\Sigma =
64.1$\,stars\,pc$^{-2}$, compared to $\Sigma = 13.1$\,stars\,pc$^{-2}$
for the region average. $\Sigma_{\rm LDR} = 4.9$, and a KS test
returns a $p$-value of $8 \times 10^{-7}$ that they share the same
underlying parent distribution.   

This is exactly as expected as the set-up is such that $\Sigma_{\rm
  LDR}$ should find mass segregation by its definition of it.

\subsubsection{Group segregation ratio, $\Omega_{\rm GSR}$}

\begin{figure*}
  \begin{center}
\setlength{\subfigcapskip}{10pt}
\hspace*{-0.3cm} \subfigure[Groups identified by MST $d_{\rm
    break}$]{\label{kirk_omega_hsd-c}\rotatebox{270}{\includegraphics[scale=0.4]{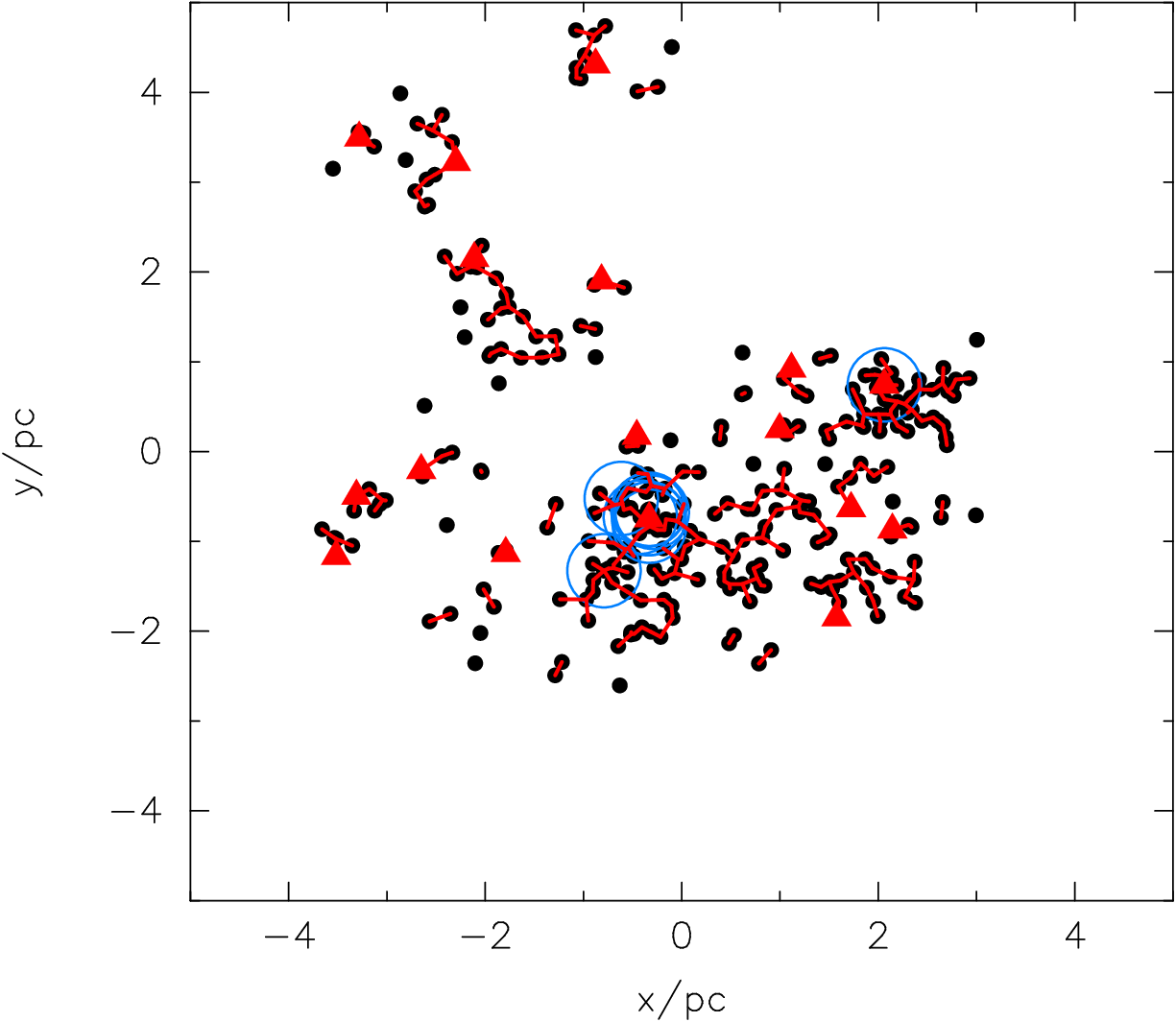}}}  
\hspace*{0.5cm} \subfigure[Mass ratio versus position offset ratio for
  groups]{\label{kirk_omega_hsd-d}\rotatebox{270}{\includegraphics[scale=0.4]{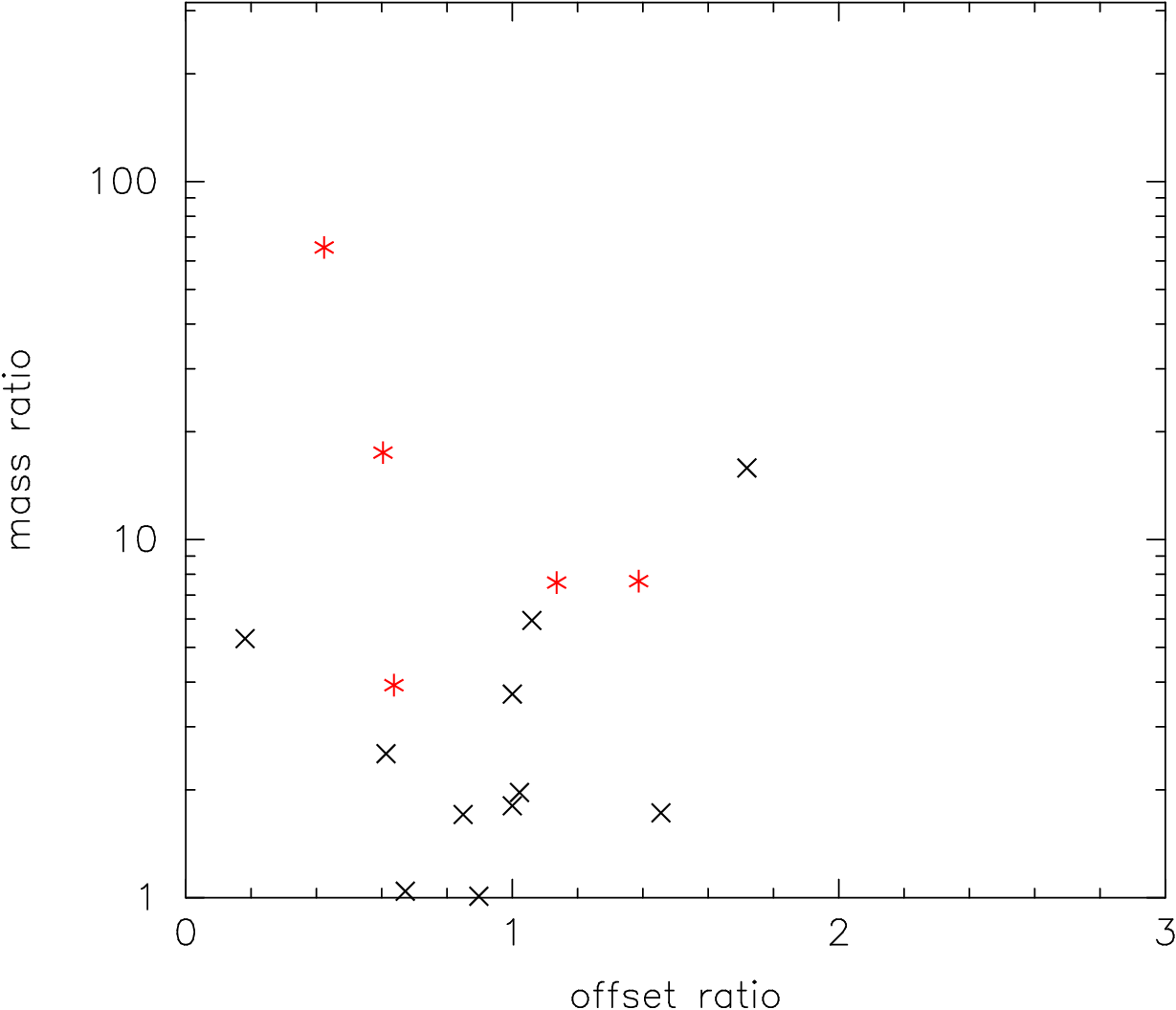}}}
\end{center}
  \caption[bf]{Mass segregation as defined by the $\Omega_{\rm GSR}$
    method for the stellar distribution shown in Fig.~\ref{map_hsd}
    where the most massive stars are in areas of highest stellar
    density. The stellar groups are identified as shown in
    Fig.~\ref{kirk_omega_ran-a} and Fig.~\ref{kirk_omega_ran-b}. In
    panel (a) we show the groups identified using MST $d_{\rm
      break}$. In the groups in which there are 3 or more stars the
    most massive star in the group is shown by the solid red
    triangle. The positions of the ten most massive stars in the
    \emph{full distribution} are shown by the large open blue circles
    (these correspond to the filled red circles in
    Fig.~\ref{map_hsd}). In panel (d) we show the mass ratio of the
    most massive star in each group to the group median mass versus
    the ratio of the position of the most massive star to the median
    position of the group. Groups with ten or more stars are shown by
    the red asterisks.}
  \label{kirk_omega_hsd}
\end{figure*}

As in Section~\ref{mass:hcc} and shown in Fig.~\ref{kirk_omega_hcc},
MST $d_{\rm break}$ is the same as that calculated in
Section~\ref{mass:ran} because the spatial distribution has not
changed. The groups identified by  MST $d_{\rm break}$ are shown in
Fig.~\ref{kirk_omega_hsd-c}, and the most massive star in each group
is shown by the red triangle. The 10 most massive stars in the
distribution are shown by the blue circles. This time, none of these
10 massive stars are not in groups, but we have a significant problem that one
group contains 9 of them (and so 8 will be discarded from the
analysis).   When we determine whether that group is mass
segregated according to the \citet{Kirk10} method, we are effectively
ignoring the positions of 8 of these stars. This time, three of the
five groups with $N \geq 10$ are mass segregated, and the largest
group (containing 9 of our most massive stars in the full
distribution) is also mass segregated according to $\Lambda_{\rm MSR}$
and $\Sigma_{\rm LDR}$. However, the other large group (containing the
single massive star) is not mass segregated according to the
\citet{Kirk10} method, but it is with $\Lambda_{\rm MSR}$ and
$\Sigma_{\rm LDR}$. For the groups with $N > 2$, $\Omega_{\rm GSR} =
0.59$, and for $N \geq 10$ $\Omega_{\rm GSR} = 0.60$, i.e.\,\,more
groups than not are mass segregated according to this method. 

%\newpage

\section{Discussion}
\label{discuss}

When attempting to find `mass segregation' in a region it is
absolutely critical to clearly define what is meant by `mass
segregation'.  Confusion between apparently contradictory results for
`mass segregation' between different methods occurs because the
different methods are searching for different things.  For 
example, \citet{Maschberger11} find mass
segregation according to $\Sigma_{\rm LDR}$ in the hydrodynamical
simulations of star formation from \citet{Bonnell08}, but do not find
mass segregation with $\Lambda_{\rm MSR}$.  We contend that this is
not condradictory, rather just different definitions of `mass
segregation' \citep[e.g.][]{Parker14b}.

A definition of mass segregation based on relaxation and equipartition
in a dynamically old system is one in which {\em the most massive
  stars are closer together than expected by random chance}.  It is
this definition that is proped by radial mass function methods ($\mathcal{M}_{\rm MF}$) and
$\Lambda_{\rm MSR}$.  In searching for this type of mass segregation
$\Lambda_{\rm MSR}$ is more useful as it does not require a centre to
be defined and can deal with complex (substructured) distributions.

The $\Sigma_{\rm LDR}$ method defines mass segregation differently --
in this case mass segregation is that {\em the most massive stars 
are preferentially in regions of higher surface density than random}. Whilst this method does not measure `mass segregation' in the classical sense, it is extremely useful for probing the past dynamical history of a star-forming region, as the most massive stars sweep up retinues of low-mass stars during the two-body relaxation of initially dense ($>$100\,M$_\odot$\,pc$^{-3}$) regions \citep{Parker14b,Parker14e,Wright14}.

One way of avoiding confusion between $\Lambda_{\rm MSR}$ and $\Sigma_{\rm LDR}$ is to make the definition of mass segregation more stringent, for example that the most massive stars should be globally more concentrated \emph{and} be at the centre of individual groups. However, this requires the somewhat arbitrary definition of groups, which is arguably impossible if all the stars formed in the same star formation `event' in the same molecular cloud, and so any boundary between groups is necessarily artificial. Furthermore, as we have seen in Section~\ref{mass:hcc} a spatial distribution that few would argue is not mass segregated would fail this definition.

In this context it is unclear to the authors what 
exactly $\Omega_{\rm GSR}$ is searching for, or what definition of
`mass segregation' it involves.  We have also identified a number of
problems with the $\Omega_{\rm GSR}$ method which we feel makes it
unsuitable for finding `mass segregation'.  

Firstly, it is unclear if the group identification is in any way
finding `real' groups.

Secondly, the identification of groups discards any stars that are not
in an $N>2$ group  \citep[or higher $N$, depending on the number of stars per group as defined in a particular analysis,][]{Kirk10,Kirk14}.  This ignores many stars, removing them from
further analysis, even if they are amoung the most massive stars in
the region.

Thirdly, once groups have been identified the method only considers
the most massive star in that group, discarding information on the
masses of any other stars.

Forthly, groups are assumed to have a `centre' from which distances
can be measured, essentially performing a `radial mass function' approach based
on a single massive star in a small-$N$ subset of the total population.

Finally, it is worth noting that small-$N$ statistics can play an
important role in obtaining any information from the $\Omega_{\rm
  GSR}$ method.  In the examples we showed above there are only five
groups with $N>10$.  For no `mass segregation' we would expect 2 or 3
to show no signal, however it would not be unusual for only 0 or 1 to
show no signal, or 4 or 5 to.

Variations on this final point is important for all methods.  A
positive signal for `mass segregation' in-and-of-itself may not tell
us much.  As we saw in the example random distribution we used above
(Section~\ref{mass:ran:lambda}), $\Lambda_{\rm MSR}$ found mass segregation at 2-$\sigma$
significance.  This is a result we would expect 1-in-20 times, and
examining our ensemble of simulations we find this is indeed the case
(and some show `inverse mass segregation' in which $\Lambda_{\rm MSR}
< 1$).  This `random noise' effect has been seen in ensembles of
simulations \citep[see][]{Parker15a}.

Based on this, it is quite possible that small signatures of `mass segregation' such
as the apparently inverse mass segregation found by \citet{Parker11b}
in Taurus might well have been over-interpreted and are quite possibly
consistent with a random distribution of the most massive stars. This highlights the requirement for more than one technique to be applied to any search for mass segregation in an observed region.

\section{Conclusions}
\label{conclude}

We have experimented with four methods used to find `mass
segregation': the radial mass function method $\mathcal{M}_{\rm MF}$ \citep[e.g.][]{Sagar88,Sabbi08}, $\Lambda_{\rm MSR}$ \citep{Allison09a}, $\Sigma_{\rm LDR}$ \citep{Maschberger11}, and $\Omega_{\rm GSR}$ \citep{Kirk10}. Our results can be summarised as follows.\\

(i) Only in smooth, spherical, centrally concentrated distributions
  (e.g. Plummer spheres) do all methods find `mass segregation'.  In
  more complex, substructured distributions different methods can find
  different things because they define `mass segregation' differently.

(ii) Only $\Lambda_{\rm MSR}$ measures `classical' mass segregation
  where the massive stars are concentrated in particular regions 
  without having to define a cluster centre.

(iii) The radial mass function method $\mathcal{M}_{\rm MF}$ searches for `classical' mass
  segregation, but requires a centre to be defined and then assumes
  spherical symmetry.

(iv) $\Sigma_{\rm LDR}$ measures a different `mass segregation' where
  the massive stars are in regions of higher than average surface
  density without having to define a cluster centre.  The massive
  stars may, or may not, also be concentrated together.

(v) $\Omega_{\rm GSR}$ finds groups that may, or may not, be
  physically important and then defines a `centre'.  In doing so it
  can exclude very significant information on some of the most massive
  stars in a region (sometimes excluding them from the analysis entirely).\\

We conclude that of the methods currently in use, by far the most
useful are $\Lambda_{\rm MSR}$ and $\Sigma_{\rm LDR}$.  They use all
of the information on all of the stars in a region without assuming
anything about the spatial distributions.  We reiterate, however, that
they measure different definitions of `mass segregation' and so should
be used in tandem.

Finally, we note that marginal signals of mass segregation (as
  found by any method) in observed star-forming regions may not have
  anything to do with the physics of star formation, and any analysis
  should be accompanied by a suite of simulations of synthetic regions
  like those presented here. In a future paper, we will also examine
  the potentially significant and serious problems of analysing
  projected distributions and attempting to extract information on the
three dimensional properties.

\section*{Acknowledgements}

We thank the anonymous referee for their comments and suggestions, which greatly improved the original manuscript. RJP acknowledges support from the Royal Astronomical Society in the
form of a research fellowship.

\bibliography{general_ref}

\label{lastpage}

\end{document}